\documentclass[12pt]{article}
\usepackage{latexsym}
\usepackage{amsmath}
\usepackage{amsfonts}
\usepackage{amssymb}
\usepackage{graphicx}
\usepackage{color}
\begin{document}
\input epsf
\def\be{\begin{equation}}
\def\bea{\begin{eqnarray}}
\def\ee{\end{equation}}
\def\eea{\end{eqnarray}}
\def\d{\partial}
\definecolor{red}{rgb}{1,0,0}
\long\def\symbolfootnote[#1]#2{\begingroup%
\def\thefootnote{\fnsymbol{footnote}}\footnote[#1]{#2}\endgroup}
\renewcommand{\a}{\left( 1- \frac{2M}{r} \right)}
\newcommand{\dm}{\begin{displaymath}}
\newcommand{\edm}{\end{displaymath}}
\newcommand{\com}[2]{\ensuremath{\left[ #1,#2\right]}}
\newcommand{\la}{\lambda}
\newcommand{\eps}{\ensuremath{\epsilon}}
\newcommand{\half}{\frac{1}{2}}
\newcommand{\field}[1]{\ensuremath{\mathbb{#1}}}
\renewcommand{\l}{\ell}
\newcommand{\bl}{\left(\l\,\right)}
\newcommand{\normljk}{\langle\l,j,k|\l,j,k\rangle}
\newcommand{\N}{\mathcal{N}}
\renewcommand{\b}[1]{\mathbf{#1}}
\renewcommand{\v}{\xi}
\newcommand{\tr}{\tilde{r}}
\newcommand{\ttheta}{\tilde{\theta}}
\newcommand{\tgamma}{\tilde{\gamma}}
\newcommand{\bg}{\bar{g}}

\renewcommand{\implies}{\Rightarrow}

\newcommand{\z}{\ensuremath{\ell_{0}}}
\newcommand{\temp}{\ensuremath{\sqrt{\frac{2\z+1}{\z}}}}
\newcommand{\twomatrix}[4]{\ensuremath{\left(\begin{array}{cc} #1 & #2
\\ #3 & #4 \end{array}\right) }}
\newcommand{\columnvec}[2]{\ensuremath{\left(\begin{array}{c} #1 \\ #2
\end{array}\right) }}
\newcommand{\e}{\mbox{\textbf{e}}}
\newcommand{\gm}{\Gamma}
\newcommand{\bt}{\bar{t}}
\newcommand{\bphi}{\bar{\phi}}
\newcommand{\m}{\ensuremath{\mathbf{m}}}
\newcommand{\n}{\ensuremath{\mathbf{n}}}
\renewcommand{\theequation}{\arabic{section}.\arabic{equation}}
\newcommand{\newsection}[1]{\section{#1} \setcounter{equation}{0}}
\newcommand{\p}{p}
\newcommand{\tmu}{\tilde{\mu}}
\newcommand{\slthree}{SL(3,\mathbb{R})}
\newcommand{\khat}{\hat{k}}
\newcommand{\mut}{\tilde{\mu}}
\newcommand{\RR}[2]{\ensuremath{\mathcal{R}_{#1 #2}}}
\newcommand{\tRR}[2]{\ensuremath{\bar{\mathcal{R}}_{#1 #2}}}
\newcommand{\DD}[2]{\ensuremath{\mathcal{D}_{#1 #2}}}
\newcommand{\pp}{\ensuremath{\Delta p}}

\vspace{20mm}
\begin{center} {\LARGE Reduction without reduction} \\ \vspace{5mm}
{\large Adding KK-monopoles to five dimensional stationary axisymmetric solutions}
\\
\vspace{20mm} {\bf Jon Ford, Stefano Giusto, Amanda Peet and Ashish Saxena}\\ 
\vspace{2mm}
\symbolfootnote[0]{ {\tt jford@physics.utoronto.ca,  giusto@physics.utoronto.ca, amanda.peet@utoronto.ca,  ashish@physics.utoronto.ca}} 
Department of Physics,\\ University of Toronto,\\ 
Toronto, Ontario, Canada M5S 1A7.\\
\vspace{4mm}
\end{center}
\vspace{10mm}
\begin{abstract}
We present a general method to add KK-monopole charge to any asymptotically flat stationary axisymmetric solution of five dimensional General Relativity. The technique exploits the underlying SL(3,R) invariance of the system by identifying a particular element of the symmetry group which changes the asymptotic boundary condition and adds KK-monopole charge. Furthermore, we develop a set of technical tools which allow us to apply the SL(3,R) transformations to solutions produced by the Inverse Scattering method. As an example of our methods, we construct the exact solution describing a static black ring carrying KK-monopole charge. 
\end{abstract}

\thispagestyle{empty}
\newpage
\setcounter{page}{1}
\newsection{Introduction}\label{sectone}
The KK-monopole is a remarkable exact solution of the five dimensional Kaluza-Klein theory~\cite{grossperry}. It describes, from a four dimensional perspective, the geometry of a magnetic monopole. From a five dimensional point of view, the geometry is completely regular and locally asymptotic to $\mathbb{R}^{3,1}\times S^1$. By a slight abuse of terminology, we will refer to solutions with such asymptotics as ``four-dimensional''.\footnote{The justification for this terminology is that if one reduces on the $S^{1}$ the resulting Einstein frame metric asymptotically approaches four dimensional Minkowski space. Of course, to be truly ``four-dimensional'' one would require the resulting dilaton and the Kaluza-Klein gauge field in four dimensions to be trivial.} The remarkable properties of the monopole solution arise from the fact that at distances much smaller than the radius of the KK direction, $R_{K}$, the geometry becomes isometric to five dimensional Minkowski space. Indeed, the radius of the KK direction can be thought of as a modulus which can be tuned to interpolate between four-dimensional ($R_{K}\rightarrow 0$) and five dimensional ($R_{K}\rightarrow \infty $) geometries. 

This property has been put to good use in the context of supergravity theories. In that case, given a supersymmetric solution in five dimensions, one can, in a systematic manner add KK-monopole charge to the solution and thereby interpolate between the original five-dimensional solution and a new four-dimensional solution.  This has been possible due to the classification of all supersymmetric solutions of certain supergravity theories in five dimensions~\cite{hullpakis}. Moreover, in the supersymmetric case one expects that appropriately defined partition functions associated to the microscopic systems dual to the gravity solutions carrying the KK-monopole charge do not depend on the modulus $R_{K}$. This has led to a remarkable connection between four and five-dimensional black hole partition functions~\cite{gsy}. From the gravity point of view, the essential simplifications come from supersymmetry and assumptions on the existence of further isometries of the solution. Under appropriate conditions, the solution is then determined by a set of harmonic functions on three dimensional flat space. In this circumstance, given an asymptotically flat five dimensional solution, the problem of adding KK-monopole charge to it reduces to adding some constants to these harmonic functions. This construction has been utilized both to find new supersymmetric multi-black hole configurations~\cite{eemr,bkw} and to construct smooth geometries representing microstates for four-dimensional black holes~\cite{bk,ours4charge,gimon,myunpub}.

For non-supersymmetric solutions, the harmonic function structure breaks down and one lacks a systematic procedure to add KK-charge~\cite{glr}. Of course, if one starts from a four dimensional solution i.e. with $\mathbb{R}^{3,1}$ asymptotics (e.g. the Kerr solution) there is a straightforward way to add KK-charge, namely by adding a trivial KK circle, boosting along it and then performing dualities to convert the resulting momentum  to a KK-monopole charge. A more sophisticated approach to doing this would be to use the $\slthree$ symmetry of Einstein gravity in five dimensions, discovered by Maison in~\cite{maison}, to add not only KK-monopole charge but also KK-electric charge. This $\slthree$ symmetry represents the duality group of pure gravity in five dimensions restricted to stationary geometries possessing at least one space-like Killing vector. An $SO(2,1)$ subgroup of this $\slthree$ has the property that it preserves $\mathbb{R}^{3,1}\times S^1$ asymptotics and thus sends the set of four-dimensional geometries into itself. It is this $SO(2,1)$ subgroup that was used in \cite{rasheed, larsen} to add KK magnetic and electric charges to the Kerr metric. The resulting solution has the property that if one looks at the geometry at distances much smaller than the radius of the KK direction one recovers the five-dimensional Myers-Perry metric. One could thus ask the question if it would be possible to generate the same solution by starting directly with the Myers-Perry metric and placing it at the tip of a KK-monopole geometry. More generally one could envisage a technique that would place any five-dimensional asymptotically flat geometry at the tip of a KK-monopole, similarly to what can be done for supersymmetric five-dimensional solutions. The main result of this paper is to provide such a technique for stationary axisymmetric five-dimensional solutions.

It was shown in \cite{gs} that the $\slthree$ symmetry mentioned above can be extended to act non-trivially on asymptotically $\mathbb{R}^{4,1}$ solutions. In this paper we show that one particular element of the $\slthree$, which we refer to as $D$, has the property that it converts solutions with $\mathbb{R}^{4,1}$ asymptotics into solutions with $\mathbb{R}^{3,1}\times S^1$ asymptotics, and in the process adds KK-monopole charge to the solution. In this sense, $D$ is the operator that realizes the 4D-5D connection for solutions in pure gravity. 

In this reduction, one combination of the two angular momenta of the five-dimensional solution is transformed into the KK-electric charge of the four-dimensional solution and the orthogonal combination becomes the four-dimensional angular momentum. This construction makes manifest a symmetry that is not evident from a four-dimensional point of view. Indeed, the starting five-dimensional solution has an obvious symmetry that exchanges the two five-dimensional angular momenta. This operation was referred to as the ``flip'' in~\cite{gs}. Once the five-dimensional system has been reduced to four dimensions after the addition of KK-charge, this symmetry exchanges the KK-momentum charge with the four-dimensional angular momentum. This observation was used in~\cite{horowitzkerr} to relate the D0-D6 system with no rotation to the extremal Kerr solution in a KK-monopole geometry. 

Of course, in the non-supersymmetric case one does not expect that the partition functions associated to the gravity solutions would be independent of the KK radius, in general. Thus in this case one cannot expect an identity between the 4D and 5D partition functions. However it might be possible to derive stronger results for extremal solutions. One example of this connection for extremal solutions has been established in \cite{emparanhorowitz,emparanmaccarrone}, where the entropy of the extremal Myers-Perry black hole has been given a microscopic derivation. An interesting open problem would be to provide a similar explanation for the entropy of the vacuum black ring of \cite{emparanreall2,pomeransky}. For this, it would be useful to construct a black ring solution carrying KK magnetic and electric charges. The techniques developed in this paper should, in principle, allow one to do this. As a first step, we construct in this paper a static black ring in a KK-monopole geometry .

The plan for the rest of the article is as follows. We start Section  \ref{secttwo} with a brief review  of the Maison formalism as applied to five dimensional stationary axisymmetric solutions. We then present the $\slthree$ matrix $D$ which toggles between four and five dimensional boundary conditions as defined below. We demonstrate the action of $D$ by showing that it maps 5D Minkowski space to the KK-monopole. In Section  \ref{sectthree}, we study in detail this action and its effect on the rod structure of static axisymmetric solutions. We find that if there are finite space-like rods present, the rod strucure is non-trivially rotated in the process of adding KK-charge. This means that even if one wants to construct static solutions with KK-charge one has to start with related non-static solutions and apply the reduction procedure to them. In Section  \ref{sectfour} we present a method using the Belinski-Zakharov inverse scattering technique \cite{bz} combined with the Maison formalism, which can be used to construct the necessary non-static seed geometries in a form that is well suited to applying the $D$ transformation. In Section  \ref{sectfive} we provide an explicit example of the solution generating procedure by using it to add KK-monopole charge to the static black ring. 

In Appendix A we show that for a 4D solution obtained using $D$, the limit in which the size of the KK-monopole is taken to be infinite reproduces the original 5D solution. In Appendix B we give the technical details associated with finding the rod orientation of the space-like rods after applying a $D$ transformation. Finally, in Appendix C we give the explicit results for the static black ring in Taub-NUT space.  

\newsection{Reviewing the $\slthree$ action}\label{secttwo}
We start with a review of the Maison formalism. For a more detailed review we refer the reader to~\cite{maison,gs}. Consider a stationary solution of five dimensional Einstein gravity with a space-like Killing vector 
${\partial\over \partial \xi^1}$. The solution can be written in the form
\be
ds^2_5 = \lambda_{ab} (d\xi^a+{\omega^a}_i dx^i)(d\xi^b+{\omega^b}_j dx^j)+{1\over \tau} ds^2_3 
\label{dec}
\ee
where $a,b=0,1$ and $\xi^0\equiv t$. $ds^2_3$ is a metric on the 3D space with coordinates $x^i$ ($i=1,2,3$); $\lambda_{ab}$ and $\omega^a_i dx^i$ are functions and 1-forms on this space, and we have defined
\be
\tau=-\mathrm{det}\lambda_{ab}
\ee 
The 1-forms $\omega^a$ can be dualized to scalars, $V_a$, as
\be
d V_a = -\tau \lambda_{ab} *_3 d\omega^b
\label{dual}
\ee
where $*_3$ is performed with the metric $ds^2_3$. As shown in \cite{maison}, the integrability of this equation is guaranteed by the Einstein equations for the metric in Eq. (\ref{dec}). Eq. (\ref{dual}) defines $V_a$ up to arbitrary constants that can be fixed by imposing some natural boundary conditions at asymptotic infinity.
The set of scalars $\lambda_{ab}$ and 
$V_a$ can be organized in the following $3\times 3$ symmetric unimodular matrix
\be
\chi=\begin{pmatrix}
\lambda_{ab} -{1\over \tau} V_a V_b & {1\over \tau} V_a\cr {1\over \tau} V_b & -{1\over \tau}
\end{pmatrix}
\ee
In terms of the matrix $\chi$, the equations of motions can be written in the compact form
\bea
&&d *_3(\chi^{-1} d\chi)=0 \label{chieq}\\
&&R^{(3)}_{ij}={1\over 4}\mathrm{Tr} (\chi^{-1}\partial_i \chi\,\chi^{-1}\partial_j \chi)
\label{eom}
\eea
where $R^{(3)}_{ij}$ is the Ricci tensor for the metric $ds^2_3$. Interpreting Eq. (\ref{chieq}) as an integrability condition we define
\be
\chi^{-1} d\chi = *_3 d\kappa
\label{kappadef}
\ee
$\kappa$ is defined up to the addition of a matrix of closed 1-forms. As shown in~\cite{gs}, $\kappa$ is related to the gauge fields $\omega^a$ by 
\be
\omega^0= -{\kappa_0}_2\,,\quad \omega^1 =-{\kappa_1}_2
\label{kappaomega}
\ee
Rewriting the Einstein equations in terms of the matrix $\chi$ has the advantage of making  the classical symmetries  
of the system manifest. Indeed, consider the following transformation
\bea
\chi\to \chi'=N\chi N^T,\quad \kappa\to \kappa'=(N^T)^{-1}\kappa N^T ,\quad ds^2_3\to ds^2_3\quad \mathrm{with}\quad N\in SL(3,\mathbb{R})
\label{5Dtrans0}
\eea
This transformation preserves the fact that $\chi$ is symmetric and unimodular and leaves the equations (\ref{chieq})-(\ref{kappadef})
invariant. Thus, given a five dimensional solution corresponding to the set of data $(\chi,ds^2_3)$, 
the geometry corresponding to $(\chi',ds^2_3)$ is also a solution of the Einstein equations.

An essential property of this $\slthree$ action is that, in general, it will not preserve the asymptotic structure of the metric. 
To see this, consider a solution which asymptotically approaches $\mathbb{R}^{3,1} \times S^{1}$ i.e. a solution of the Kaluza-Klein gravity. The asymptotic behaviour of the five dimensional metric is assumed to be described by 
\be
ds^2 = -dt^2 + dr^2 +r^2 (d\theta^2+ \sin^2\theta d\phi^2) + (dx^5)^2
\ee
Here $x^5$ parametrizes an $S^{1}$ of fixed radius. The $\chi$ for this metric (with the choice $\xi^1 = x^5$) is given by
\be
\eta_4  \equiv \begin{pmatrix}-1&0&0\cr 0&1&0\cr0&0&-1\end{pmatrix}
\ee 
This indicates that for a general asymptotically $\mathbb{R}^{3,1} \times S^{1}$  solution, the asymptotic behaviour of $\chi$ is
\be
 \chi\rightarrow \eta_4 \label{asympchi4}
\ee 
We will refer to a solution with $\chi$ approaching $\eta_4$ as `four dimensional'.

The next interesting case is that of asymptotically Minkowski boundary conditions. As an example consider five dimensional Minkowski space
\be
ds^2 = -dt^2 + dr^2 + r^2 (d\theta^2 + \sin^2\theta d\phi^2 + \cos^2\theta d\psi^2) 
\label{5dflat}
\ee
and choose $\xi^{0} =t,\ \xi^{1} =\ell(\psi+\phi)$ where $\ell$ is some arbitrary length scale. With this choice, $\chi$ becomes constant asymptotically.
\be
\chi= \begin{pmatrix}-1&0&0\cr0&0&1\cr0&1& -\frac{4\ell^2}{r^2}
\end{pmatrix} \stackrel{r\rightarrow \infty}{\longrightarrow} \eta_5\equiv \begin{pmatrix}-1&0&0\cr0&0&1\cr0&1&0
\end{pmatrix}
\ee
Thus, an asymptotically $\mathbb{R}^{4,1}$ solution will have a $\chi$ matrix approaching $\eta_{5}$ as $r\rightarrow \infty$ and we will refer to such solutions as `five-dimensional'. 

We have seen that for four as well as five dimensional solutions, the matrix $\chi$ is asymptotically constant. This leads to the possibility that there might be a (constant) $\slthree$ transformation which would convert a five dimensional solution to a four dimensional one and vice versa. Indeed, such a transformation was presented in~\cite{gs}. The matrix $\eta_5$ is related to $\eta_4$ by an $\slthree$ matrix $D$ 
\be
\eta_4 = D\eta_5 D^{T}\,, \quad D=\begin{pmatrix}1&0&0\cr 0& {1\over \sqrt{2}}& {1\over \sqrt{2}}\cr  0& -{1\over \sqrt{2}}& {1\over \sqrt{2}}
\end{pmatrix}
\ee
An interesting consequence of the existence of this transformation is that given a five dimensional vacuum solution, one can convert it into a KK solution by applying $D$. Roughly speaking, the action of $D$ is to put the given five dimensional vacuum solution at the center of the KK-monopole, and in the process change the asymptotics to $\mathbb{R}^{3,1}\times S^1$. As the simplest example of this phenomena, one can consider the five dimensional Minkowski space itself. Based on the intuition above, one would expect to recover the KK-monopole by applying $D$. This is easiest to see if one employs the Gibbons-Hawking form~\cite{gibbonshawking} for the four dimensional spatial sections of $\mathbb{R}^{4,1}$. Starting from Eq. (\ref{5dflat}) and performing the following coordinate changes
\be
\xi^{1} =\ell(\psi+\phi),\ \tilde{\phi} = \psi-\phi,\ \rho=\frac{r^2}{4\ell},\ \tilde{\theta}= 2\theta
\ee
brings the metric to the following form
\be
ds^2_5= -dt^2 + V^{-1}  (d\xi^{1} +\ell \cos\tilde{\theta} d\tilde{\phi})^2 + \frac{1}{V} (d\rho^2+\rho^2d\tilde{\theta}^2+\rho^2 \sin^2\tilde{\theta} d\tilde{\phi}^2),\ V= \frac{\ell}{\rho} 
\ee 
In this form, it is easy to read off the vector potentials $\omega^a$ and the three-dimensional base space. As the three dimensional metric turns out to be flat, it is also easy to find the twist potentials defined in Eq. (\ref{dual}) and hence the $\chi$ and $\kappa$. We find
\be
\chi = \begin{pmatrix}-1&0&0\cr0&0&1\cr0&1& -V
\end{pmatrix},\ \kappa = \begin{pmatrix}0&0&0\cr0&0&-\ell \cos\tilde{\theta}\cr0&0& 0
\end{pmatrix} d\tilde{\phi}
\ee
Applying $D$ to the above data and reconstructing the metric we find
\be
ds^2= -dt^2 + \tilde{V}^{-1}  (d\xi^{1} +\frac{1}{2}\ell \cos\tilde{\theta} d\tilde{\phi})^2 + \frac{1}{\tilde{V}} (d\rho^2+\rho^2d\tilde{\theta}^2+\rho^2 \sin^2\tilde{\theta} d\tilde{\phi}^2),\ V= 1+\frac{\ell}{2\rho} 
\ee 
which is isometric to the Gross-Perry \cite{grossperry} metric with charge $Q={\ell}/{2}$. 

To summarize, we have shown that applying $D$ to five-dimensional flat space effectively adds KK-charge to it and changes its asymptotics appropriately. It is natural to extend this action to other asymptotically flat five dimensional vacuum solutions to add KK magnetic charge to them. This is the subject of the following sections. An interesting feature of adding KK-monopole charge to five dimensional solutions is that setting the charge to zero does not lead to the starting solution. This is well known for the case of the KK-monopole itself, where in order to recover the five dimensional Minkowski metric (with discrete identifications in general) is also necessary to take the limit $\ell\rightarrow \infty$. In Appendix A we generalise this result and show that the above procedure of taking $\ell\rightarrow \infty$ recovers the starting five dimensional solution in the general case as well.

\newsection{Action of $D$ on static solutions}\label{sectthree}
All static axisymmetric solutions of five-dimensional general relativity can be constructed explicitly, using the results of \cite{emparanreall}. Here we briefly review the general solution of this type, with $\mathbb{R}^{4,1}$ asymptotic boundary conditions; further details can be found in \cite{gs}. We will then analyze the solution obtained by application of the $SL(3,\mathbb{R})$ rotation
$D$ on these static solutions.

\subsection{The setup}\label{sectthreeone}
Static axisymmetric solutions of $d=5$ GR are completely characterized by their rod structure, defined in \cite{emparanreall}. Consider a solution with $N$ finite rods: the $i$-th rod starts at $z=p_i$ and ends at $z=p_{i+1}$; we denote by $L_i$ its length, $L_i=p_{i+1}-p_i$. In addition the solution has two semi-infinite rods, extending along $(-\infty  ,  p_<)$ and $(p_>  , +\infty)$. By construction
\be
p_1=p_<,\quad p_i = \sum_{j=1}^{i-1} L_j , \quad p_> = p_< +\sum_{j=1}^N L_j
\ee 
In five dimensions, the stationary axisymmetric solutions that we consider\footnote{More general black hole solutions with only one
axial symmetry are conjectured to exist \cite{reall, hiw}, but none of them is explicitly known yet.} have three Killing vectors, associated with the
time coordinate $t$ and the two azimuthal angles $\phi$ and $\psi$.  For a static solution, the eigenvectors associated to each rod have only one non-zero component, either along $t$, or along 
$\phi$ or $\psi$. We denote with the index $i_0, i_1$ and $i_2$ the set of rods whose eigenvectors point along $t,\phi$ and $\psi$, respectively.

In Weyl's canonical coordinates, the metric has the form
\be
ds^2 = -e^{2 U_0} dt^2 + e^{2 U_1} d\phi^2 + e^{2 U_2} d\psi^2 + e^{2 \nu}(dr^2+dz^2)
\label{static}
\ee
where $U_I$ ($I=0,1,2$) are harmonic functions on $\mathbb{R}^3$, satisfying the constraint
\be
U_0+U_1+U_2 = \log r
\ee 
They can be most conveniently expressed in terms of the combinations
\be
\mu_i = \sqrt{r^2 + (z-p_i)^2}-(z-p_i)
\ee
The harmonic functions are given by\footnote{
Of course, the sums over $i_0$, $i_1$ and $i_2$ are different, because those variables are defined as indexing rods associated with the different directions $t$, $\phi$ and $\psi$.
}

\bea
U_0=\sum_{i_0} U_{i_0} , \quad U_1 = U_>+\sum_{i_1} U_{i_1},\quad U_2 = U_<+\sum_{i_2} U_{i_2}
\eea
where
\bea
U_i = {1\over 2}\log {\mu_i\over \mu_{i+1}}\,,\quad U_<= {1\over 2}\log {r^2\over \mu_<},\quad U_> = {1\over 2}\log \mu_> 
\eea
The function $\nu$ can be derived from the $U_I$'s, by inverting the differential relations
\be
\partial_r \nu = -{1\over 2 r}+{r\over 2} \sum_{I=0}^2[(\partial_r U_I)^2 -(\partial_z U)^2 ]\,,\quad \partial_z \nu = r \sum_{I=0}^2\partial_r U_I \partial_z U_I
\ee

The solution (\ref{static}) goes over to $\mathbb{R}^{4,1}$ at asymptotic infinity. As explained in the previous section , we can generate from it a solution carrying KK-monopole charge, which asymptotes to
$\mathbb{R}^{3,1}\times S^1$. Let the direction of $S^1$ be
\be
\xi^1=\ell (\psi+\phi)
\ee
Let us also denote the azimuthal coordinate of $\mathbb{R}^{3,1}$ as
\be
\tilde\phi = \psi-\phi
\ee
To apply $SL(3,\mathbb{R})$ transformations one has to rewrite the metric (\ref{static}) in the form (\ref{dec}) and compute 
the corresponding $\chi$ and $\kappa$ matrices. The metric (\ref{static}) can be brought to the form (\ref{dec}) with
\bea
&&\lambda_{00}=-e^{2 U_0} ,\quad \lambda_{11}={e^{2 U_1}+e^{2 U_2}\over 4 \ell^2}, \quad \lambda_{01}=0, \quad \omega^0=0, \quad \omega^1 = -\ell {e^{2 U_1} - e^{2 U_2} \over e^{2 U_1} + e^{2 U_2}}d
\tilde\phi\nonumber\\
&& ds^2_3 = {r^2\over 4 \ell^2}d{\tilde \phi}^2 +\tau e^{2\nu}(dr^2+dz^2), \quad \tau= e^{2 U_0}{e^{2 U_1}+e^{2 U_2}\over 4 \ell^2}
\eea 
The potentials $V_0$ and $V_1$ are given by
\bea
V_0 &=& 0\\
V_1 &=& {1\over 8\ell^2}\Bigl[{(r^2+\mu_<\mu_>)(\mu_<+\mu_>)\over \mu_<\mu_>}\\ \nonumber
&&+\sum_{i_1} {(r^2-\mu_{i_1+1}\mu_{i_1})(\mu_{i_1+1}-\mu_{i_1})\over \mu_{i_1+1}\mu_{i_1}} - \sum_{i_2} {(r^2-\mu_{i_2+1}\mu_{i_2})(\mu_{i_2+1}-\mu_{i_2})\over \mu_{i_2+1}\mu_{i_2}} \Bigr]\nonumber
\eea
The matrix $\chi$ can be straightforwardly constructed from the quantities given above
\be
\chi=\begin{pmatrix}-e^{2 U_0}&0&0\cr 0& {e^{2 U_1}+e^{2 U_2}\over 4 \ell^2}- {4\ell^2 V_1^2\over e^{2 U_0}(e^{2 U_1}+e^{2 U_2})}& {4\ell^2 V_1\over e^{2 U_0}(e^{2 U_1}+e^{2 U_2})}\cr 0&  {4\ell^2 V_1\over e^{2 U_0}(e^{2 U_1}+e^{2 U_2})}& - {4\ell^2 \over e^{2 U_0}(e^{2 U_1}+e^{2 U_2})}\end{pmatrix}
\label{chistatic}
\ee 
The matrix $\kappa$ is given by
\be
\kappa = \begin{pmatrix}-\tilde U_0&0&0\cr 0& \omega^1 V_1 -\tilde U_+  & -\omega^1\cr
0& \omega^1 V_1^2 +\tilde V_1 & -\omega^1 V_1 +{\tilde U}_0+ {\tilde U}_+\end{pmatrix}d\tilde\phi
\label{kappastatic}
\ee 
where
\bea
{\tilde U}_0 \!\!&=&\!\! {1\over 4\ell} \sum_{i_0} {(r^2-\mu_{i_0+1}\mu_{i_0})(\mu_{i_0+1}-\mu_{i_0})\over \mu_{i_0+1}\mu_{i_0}}\\
{\tilde U}_+\!\!&=&\!\!{1\over 4\ell}\Bigl[{r^2-\mu_<\mu_>\over \mu_>}+\sum_{i_1}{r^2 (\mu_{i_1+1}-\mu_{i_1})\over \mu_{i_1+1}\mu_{i_1}} -\sum_{i_2} (\mu_{i_2+1}-\mu_{i_2})\Bigr]\\\nonumber
{\tilde V}_1 \!\!&=&\!\! -{1\over 32\ell^3}\Bigl[{(r^4-\mu^2_<\mu^2_>)(\mu^2_<+\mu^2_>)\over \mu^2_<\mu^2_>}\nonumber\\
&&+\sum_{i_1} {(r^4+\mu_{i_1+1}^2\mu_{i_1}^2)(\mu^2_{i_1+1}-\mu^2_{i_1})\over \mu^2_{i_1+1}\mu^2_{i_1}} -\sum_{i_2} {(r^4+\mu_{i_2+1}^2\mu_{i_2}^2)(\mu^2_{i_2+1}-\mu^2_{i_2})\over \mu^2_{i_2+1}\mu^2_{i_2} } \Bigr]
\nonumber
\eea
Arbitrary additive constants have been fixed by demanding that every component of $\kappa$ behaves at infinity as $z/\sqrt{r^2+z^2} d{\tilde \phi}$.
This condition is, in turn, required by the asymptotics and the definition (\ref{kappadef}) of $\kappa$.

\subsection{Action of $D$ on the rod structure}\label{sectthreetwo}

Consider now the metric obtained by applying the $\slthree$ transformation $D$ to the $\chi$ and $\kappa$ above
\be
{\bar \chi}=D\chi D^T, \quad {\bar \kappa}=D\kappa D^T,\quad d{\bar s}^2_3=ds^2_3
\label{afterD}
\ee
This metric does not fall into the category of static solutions; however, the concept of rod structure
can be generalized to any stationary solution, as shown in the seminal work of \cite{harmark}. We will
study the rod structure for the metric (\ref{afterD}) below. 

For this purpose, let us look at the $r\to 0 $ limit of the metric described by $\bar \chi$ and $\bar \kappa$. Let $\bar \lambda$, 
$\bar\omega^a$, $\bar\tau$ be the metric coefficients derived from $\bar\chi$ and $\bar\kappa$. 

As explained in
\cite{gs}, there are two distinct cases to consider. If, for $r\to 0$ and some value of $z$, 
$\bar\tau$ vanishes as $r^2$, then $z$ lies inside a time-like rod, whose corresponding eigenvector is given by (in a vector basis given by $({\partial\over \partial t}, {\partial\over \partial \xi^1},{\partial\over \partial \tilde\phi})$)
\be
v=(v_0^0,v_0^1,0)
\ee 
where $(v_0^0,v_0^1)$ is in the kernel of the matrix $\bar\lambda$ (this kernel has to be non-empty because
$\mathrm{det}\bar\lambda=-\bar\tau=0$).
If $\bar\tau$ remains finite as $r\to 0$, then one is inside a space-like rod, and the corresponding
eigenvector is
\be
v=(-\bar\omega^0_{\tilde\phi}, -\bar\omega^1_{\tilde\phi},1)
\label{spaceeigen}
\ee 
where $\bar\omega^a_{\tilde\phi}$ are the components of $\bar\omega^a$ evaluated as $r\to 0$ and $z$ inside the rod under consideration.

Let us start from the first case: suppose that $z$ belongs to a time-like rod of the starting static metric (\ref{static}), i.e. $z\in(p_{i_0},p_{i_0+1})$. Then it easily follows from (\ref{chistatic}), that
for $r\to 0$, $\chi$ has the form
\be
\chi\approx {1\over f(z)} \begin{pmatrix}
-r^2 f(z) g(z)  &0&0\cr 0&{1\over \ell^2}\Bigl({f(z)^2\over g(z)}-{a^2\over r^2}\Bigr)&{a\over r^2} \cr 0&{a\over r^2}& -{\ell^2\over r^2}
\end{pmatrix}
\ee
where $f(z)$, $g(z)$ are positive functions and $a$ is a constant. After a $D$ rotation, one finds 
\bea
&&\bar\lambda\approx \begin{pmatrix}-g(z) r^2 &0\cr 0& {2 f(z)\over \ell^2 (1+a/\ell^2)^2 g(z)} \end{pmatrix}\nonumber\\
&&\bar\tau\approx {2 f(z)\over \ell^2 (1+a/\ell^2)^2} r^2
\eea
The fact that $\bar\tau\sim r^2$ implies that the present range of $z$ lies inside a time-like rod of the $D$-transformed metric. Moreover, the kernel of $\bar\lambda$ at $r=0$ is generated by the vector $(1,0)$, and thus the eigenvector associated to this time-like rod is $(1,0,0)$, the same as for the original static metric.
We thus see that the transformation $D$ does not rotate the eigenvectors of time-like rods.

Let us now consider the case in which $z$ belongs to one of the space-like rods of the starting solution 
(\ref{static}), and thus $\tau$ goes to some non-vanishing function of $z$ when $r\to 0$. It is easy to see, by explicit inspection, that the $D$-transformed metric also has the property that $\bar\tau$ does not vanish
for small $r$, and thus $D$ sends space-like rods into space-like rods. We would like to analyze how the eigenvectors associated to these rods transform under $D$. Using (\ref{spaceeigen}) for the spacelike rod eigenvector, this requires
knowledge of $\bar\omega^a_{\tilde\phi} = - (\bar\kappa_{\tilde\phi})_{a2}$. We will see that, for $r\to 0$ and
$z$ inside a space-like rod, $\kappa$ goes to a constant matrix of the form
\be
\kappa = \begin{pmatrix}{1\over \ell}k_{00}&0&0\cr 0& {1\over \ell}k_{11}& \ell k_{12}\cr 0& {1\over \ell^3}k_{21}& {1\over \ell}k_{22}\end{pmatrix} d\tilde\phi
\label{kapparod}
\ee
and thus its $D$-transformed $\bar\kappa$ has
\be
 (\bar\kappa_{\tilde\phi})_{02}=0,\quad  (\bar\kappa_{\tilde\phi})_{12}={\ell\over 2}\Bigl(k_{12}+{k_{22}-k_{11}\over \ell^2}-{k_{21}\over \ell^4}\Bigr)
\ee
Thus the eigenvector associated to a space-like rod of the $D$-transformed metric is
\be
v=\Bigl(0,{\ell\over 2}\Bigl(k_{12}+{k_{22}-k_{11}\over \ell^2}-{k_{21}\over \ell^4}\Bigr),1)
\ee
All we need to do is compute the constants $k_{ij}$; their value can be computed from the form of $\kappa$ given in (\ref{kappastatic}), and depends on which space-like rod we are considering.  We will list the explicit expressions for these constants in Appendix B. The results can be summarized as
follows. The eigenvectors $v_<$ and $v_>$ associated to the left and right semi-infinite rods have the form
\be
v_< = (0,Q,1),\quad v_>=(0,-Q,1)
\ee
where
\be
Q= {\ell\over 2}\Bigl(1-{p_>^2+p_<^2-\sum_{i_1}(p_{i_1+1}^2-p_{i_1}^2)+\sum_{i_2}(p_{i_2+1}^2-p_{i_2}^2)\over 8 \ell^4}+{c^2\over \ell^4}\Bigr)
\label{Q}
\ee
and
\be
c= {1\over 4}[\p_<+\p_>-\sum_{i_1} L_{i_1} + \sum_{i_2} L_{i_2}]
\label{c}
\ee

The eigenvector $v_i$ associated to the $i$-th finite space-like rod is
\be
v_i = (0,Q_i,1)
\ee 
where $Q_i$ is a constant that depends on the rod under consideration, and is, in general, different from
$\pm Q$. This implies that, in general, the transformation $D$ changes the {\it relative} orientation between the eigenvectors associated to space-like rods.

\subsection{Asymptotic charges}\label{sectthreethree}

In this subsection we will analyze the large distance limit of the $D$-transformed solution, and compute its mass and KK-monopole charge. One should start by computing the asymptotic behavior of static solutions (\ref{static}). This has already been worked out in \cite{gs}, and we recall those results here.

Let us define the coordinates $\rho$ and $\theta$ as
\be
r={\rho^2\over 2}\sin2\theta\,,\quad z={\rho^2\over 2}\cos2\theta
\label{asympcoord}
\ee
and introduce the parameters $\delta_I$ and $\epsilon_I$, that characterize the rod structure of the solution (\ref{static}) 
\bea
\delta_I &=& \sum_{i_I} [\p_{i_I+1}- \p_{i_I}]-\p_> \delta_{I,1}+\p_<\delta_{I,2}\\
\epsilon_I&=& \sum_{i_I} [\p_{i_I+1}^2- \p_{i_I}^2]-\p^2_> \delta_{I,1}+\p^2_<\delta_{I,2}\\
\eea
It is also convenient, because it simplifies some of the following equations, to pick the origin on the 
$z$ axis such that the metric (\ref{static}) satisfies the harmonic gauge. This is achieved by demanding
that $\delta_1=\delta_2$, which fixes the position of the left semi-infinite rod to be
\be
\p_< = -{1\over 2}\sum_{i_0}L_{i_0}-\sum_{i_2}L_{i_2}
\ee
Note that with this choice the constant $c$ in (\ref{c}) vanishes
\be
c=0
\ee

The large $\rho$ limits of the matrices $\chi$ and $\kappa$ are
\be
\chi\approx \begin{pmatrix}-1+{2\delta_0\over\rho^2}&0&0\cr 0&{\epsilon_1-\epsilon_2\over 2 \ell^2 \rho^2}& 1+{\delta_0\over \rho^2}\cr 0& 1+{\delta_0\over \rho^2}& -{4\ell^2\over \rho^2}\Bigl(1+{\delta_0\over \rho^2}\Bigr)\end{pmatrix}
\ee
\be
\kappa\approx \begin{pmatrix}-{\delta_0\over 2 \ell}&0&0\cr 0& -{\delta_1\over 2\ell}&-\ell\cr0&{\epsilon_1-\epsilon_2\over 8\ell^3}&-{\delta_1\over 2\ell}\end{pmatrix}\cos2\theta d\tilde\phi
\ee
It is now straightforward to compute the $D$-transformed matrices $\bar\chi=D\chi D^T$ and $\bar\kappa=D\kappa D^T$ and, from them, the corresponding metric. We find
\bea
ds^2&\approx& - \Bigl(1-{\delta_0\over 2 \ell \tilde r}\Bigr) dt^2+\Bigl(1-{\ell\over 2\tilde r}\Bigl(1-{\epsilon_1-\epsilon_2\over 8 \ell^4}-{\delta_0\over 2\ell^2}\Bigr)\Bigr)(d\xi^1+Q \cos\tilde\theta d\tilde\phi)^2 \nonumber\\
&&+ d{\tilde r}^2 + {\tilde r}^2 d{\tilde\theta}^2 + {\tilde r}^2 \sin^2\tilde\theta d\tilde\phi^2
\label{asymp}
\eea
where
\be
\tilde r = {\rho^2\over 4 \ell},\quad \tilde\theta = 2\theta
\ee
and
\be
Q= {\ell\over 2}\Bigl(1+{\epsilon_1-\epsilon_2\over 8 \ell^4}\Bigr)= {\ell\over 2}\Bigl(1-{p_>^2+p_<^2-\sum_{i_1}(p_{i_1+1}^2-p_{i_1}^2)+\sum_{i_2}(p_{i_2+1}^2-p_{i_2}^2)\over 8 \ell^4}\Bigr)
\ee
Note that the $Q$ defined here agrees with the one defined previously in Eq. (\ref{Q}) (recalling that with our present conventions $c=0$). 

$Q$ represents the KK-monopole charge of the system. From (\ref{asymp}) we can also read off the mass
of the system
\be
G_4 M = {\ell\over 8 }\Bigl(1-{\epsilon_1-\epsilon_2\over 8 \ell^4}+{3 \delta_0\over 2 \ell^2}\Bigr)
\ee 

\subsection{Special cases}\label{sectthreefour}

\subsubsection{Black Hole}\label{sectthreefourone}
\begin{figure}[htp]
\center
\includegraphics[totalheight=3.5 cm]{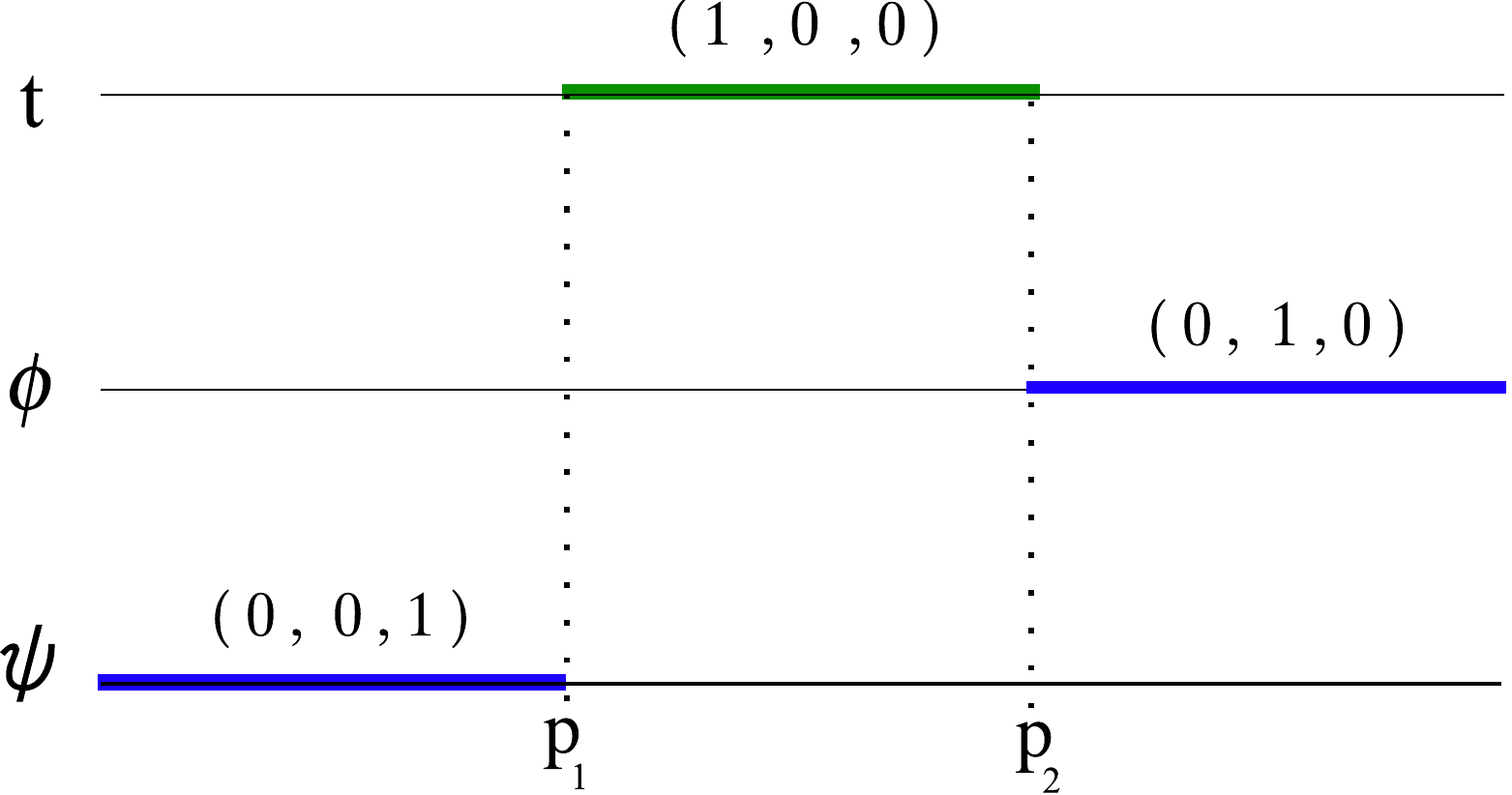}
\caption{Rod structure of the five dimensional Schwarzschild black hole.}
\end{figure}

The simplest static solution, after flat space, is the five-dimensional Schwarzschild hole. It is characterized by having a single finite
time-like rod and no finite space-like rods. Its rod structure is given in Figure 1. The rod end-points can be taken to be
\be
p_>=-p_< = k^2
\ee
The Weyl form of the static black hole metric is
\bea
ds^2 &=& -{\mu_<\over \mu_>}dt^2+ \mu_> d\phi^2 + {r^2\over \mu_<} d\psi^2 + e^{2\nu} (dr^2 + dz^2)\\
e^{2\nu}&=& \mu_> {(r^2 + \mu_< \mu_>)\over (r^2+\mu_<^2)(r^2+\mu_>^2)}
\eea
Applying the procedure described above one arrives at the following metric, representing a black hole in Taub-NUT space
\bea
ds^2 &=& -{\mu_<\over \mu_>}dt^2+{\bar\lambda}_{11}(d\xi^1 + {\bar \omega}^1)^2+ 
{1\over \bar\tau} \Bigl({r^2\over 4 \ell^2} d{\tilde\phi}^2+ \tau e^{2\nu} (dr^2+dz^2)\Bigr)\\
{\bar\lambda}_{11}&=&{8 \mu_> k^2 \ell^2\over 4 \ell^4 (\mu_>-\mu_<) + 4 k^2 \ell^2 (\mu_>+\mu_<) + k^4 (\mu_>-\mu_<)}\nonumber\\
{\bar \omega}^1 &=& {\ell\over 2}\Bigl(1-{k^4\over 4 \ell^4}\Bigr) {r^2-\mu_< \mu_>\over r^2+\mu_< \mu_>}\nonumber\\
{\bar\tau}&=&{8 \mu_< k^2 \ell^2\over 4 \ell^4 (\mu_>-\mu_<) + 4 k^2 \ell^2 (\mu_>+\mu_<) + k^4 (\mu_>-\mu_<)}\,\,,\,\, \tau = {k^2 \mu_<\over \ell^2 (\mu_>-\mu_<)}\nonumber
\label{nuthole}
\eea
This solution carries KK-monopole charge $Q$, given by
\be
Q={\ell\over 2}\Bigl(1-{k^4\over 4 \ell^4}\Bigr) 
\ee
In \cite{rasheed,larsen} a geometry carrying KK magnetic and electric charge, and angular momentum, was derived by
applying an $SO(2,1)$ transformation to the Kerr solution (times a trivial circle). It can be checked that the geometry (\ref{nuthole}) coincides with the particular case of the solution of \cite{rasheed, larsen}, in which the KK electric charge and the angular momentum are set to zero. However, we have shown above that the same geometry can be derived starting from the five-dimensional black hole rather than
the four-dimensional one. It is easy to generalise our construction so as to reproduce the general geometries of~\cite{rasheed, larsen}.
\subsubsection{Black Ring}\label{sectthreefourtwo}
The next case to consider is a static solution with two finite rods. If the solutions has to represent a black object, one of the finite rods has to be time-like. The remaining finite rod is thus space-like: up to an exchange of the coordinates $\phi$ and $\psi$, we can take this rod to be oriented along $\psi$. The full rod structure is depicted in Figure 2(a). This solution represents a static black ring, with horizon topology given by $S^2\times S^1$. It is easy to understand why this is so. Consider the region $r=0$ and $z\in[p_1,p_2]$, which constitutes the horizon. The $\psi$ circle is fibered over the interval $z\in[p_1,p_2]$ and it shrinks to zero size at the two end-points $p_1$ and $p_2$. This is topologically an $S^2$. The $\phi$ circle remains finite in this region, and represents the $S^1$ of the ring.
Note that it is crucial for the horizon topology to be $S^2\times S^1$ that the two space-like rods, at the left and the right of the time-like rod, be associated with the same direction. 

Consider now acting with the transformation $D$ on such a static solution. As we have seen in Section ~\ref{sectthreetwo}, $D$ changes the
relative orientation between finite space-like rods. This means that if we start from the configuration of Figure 2(a), in the transformed geometry the space-like rods neighboring the horizon will not be parallel, and thus the horizon topology will not be that of a ring.
Generically the solution will not even be smooth. 

To construct a ring with KK-monopole charge it is necessary that the finite space-like rod be parallel to the semi-infinite space-like rod along $\psi$ \emph{after} the application of $D$. This indicates that if we started with a solution in which this finite space-like rod was already carrying an arbitrary orientation, then one could, after applying $D$, fix this arbitrary parameter so as to force the  final orientation to be parallel to the semi-infinite rod. To be precise, if one is only interested in constructing a non-rotating black ring in Taub-NUT, one does not need to start with the finite rod in the most generic position. This is because the action of $D$, rotates the eigenvector associated to this rod only in the $\partial_{\psi}-\partial_{\phi}$ plane. As such, we need to start with the solution where this finite rod has a generic orientation in the above plane. 

Such a solution however, does not fall into the class of static solutions considered in this section. The simplest method to construct this solution is by using the well known inverse scattering transform techniques~\cite{bz,pomeranskyone} also known as the Belinski-Zakharov method. Once we have this `seed' solution we can apply $D$ to it as explained in the previous sections. However a brute force application of this strategy soon runs into a technical brick wall. This is because the solution obtained from the inverse scattering transform is complicated and solving the duality equations~(\ref{dual}) directly to find the twist potentials $V_a$ is not feasible. However, these problems can be circumvented. In the following section, we show that we can effectively ``commute'' the problem of finding the twist potentials $V_a$ and the $\chi$ and $\kappa$ matrices past the Belinski-Zakharov transformation. This means that we can algebraically relate the Maison data $(\chi,ds_3^2)$ after a BZ transform to the data before the transform. The results of the next section apply generally and can be used to construct stationary solutions and add KK-monopole charge to them without the need to solve complicated differential equations.

\newsection{Action of $D$ on solutions from Inverse Scattering}\label{sectfour}
A stationary axisymmetric solution of five-dimensional Einstein gravity can be written in Weyl form as follows
\be
ds^2 = G_{IJ} dy^I dy^J + e^{2\nu}(dr^2+dz^2)
\ee
where the coordinates $y^I$, $I=0,1,2$, are associated to the three Killing vectors of the solution; we choose $y^0=t$, $y^1=\phi$, $y^2=\psi$. The variable $r$ is defined by the condition
\be
\mathrm{det} G = -r^2
\label{Gconstraint}
\ee
Let us define the matrices
\be
U=r (\partial_r G) G^{-1}\,,\quad V=r(\partial_z G) G^{-1}
\label{uv}
\ee
In this coordinate system Einstein equations reduce to \cite{bz,harmark}
\be
\partial_r U +\partial_z V=0\label{equv}
\ee
\be
\partial_r \nu =-{1\over 2 r}+{1\over 8 r}\mathrm{Tr}(U^2-V^2)\,,\quad \partial_z\nu = {1\over 4 r} \mathrm{Tr} (U V) \label{nueq}
\ee
Eq. (\ref{equv}) is the equation that determines the matrix $G$; once $G$ is known, the differential relations (\ref{nueq}) can be solved
for $\nu$. Note that the integrability of Eq. (\ref{nueq}) is guaranteed by Eq. (\ref{equv}) and by the identity
\be
\partial_z U - \partial_r V +{1\over r} (\com{U}{ V}  + V)=0
\ee
which follows from the definitions (\ref{uv}).

Thus the problem of finding axisymmetric stationary solutions is reduced to the problem of finding solutions to Eq. (\ref{equv}).  In
\cite{bz}, Belinski and Zakharov have shown that this problem is integrable, and have given a constructive technique to produce new solutions from known ones, also known as the Inverse Scattering (IS) method. Further applications of this method have been worked out in
\cite{pomeranskyone,pomeransky,elvang,evslin}. In the following we briefly review the method and derive some new identities that are useful in order to apply $SL(3,\mathbb{R})$ transformations to solutions produced by the IS technique.

\subsection{Review of Inverse Scattering construction}\label{sectfourone}
Let $G_0$ be a known solution  to Eq. (\ref{equv}), referred to as the ``seed''. Then $G_0$ satisfies
\be
\partial_r U_0 + \partial_z V_0=0\,,\quad \mathrm{det} G_0 = - r^2
\ee
where
\be
U_0=r \partial_r G_0 G_0^{-1}\,,\quad V_0=r\partial_z G_0 G_0^{-1}
\label{equv0}
\ee
In usual applications $G_0$ is a diagonal matrix, corresponding to a static solution, but the method applies more generally to any
stationary seed metric.

One can introduce a pair of differential operators
\be
D_r = \partial_r +{2\lambda r\over r^2 +\lambda^2}\partial_\lambda\,,\quad D_z = \partial_z -{2\lambda^2 \over r^2 +\lambda^2}\partial_\lambda
\ee
and look for a $\lambda$-dependent matrix $\Psi_0(\lambda,r,z)$ satisfying
\be
D_r \Psi_0 = {r U_0 +\lambda V_0\over r^2+\lambda^2}\Psi_0\,,\quad D_z \Psi_0 = {r V_0 -\lambda U_0\over r^2+\lambda^2}\Psi_0
\label{eqpsi0}
\ee
The equations of motion for $G_0$ in Eq. (\ref{equv0}) ensure the integrability of the equation for $\Psi_0(\lambda, r,z)$.
The goal of the IS method is to construct a matrix $\Psi(\lambda,r,z)$ that satisfies the analogue of Eq. (\ref{eqpsi0})
\be
D_r \Psi = {r U +\lambda V\over r^2+\lambda^2}\Psi\,,\quad D_z \Psi = {r V -\lambda U\over r^2+\lambda^2}\Psi
\label{eqpsi}
\ee
for some new metric $G$. Note that the equation that defines $\Psi$, (\ref{eqpsi}), implies that
\be
\Psi(0,r,z) = G(r,z)
\ee
so that $G$ is known once $\Psi$ is known. 

$\Psi$ can be generated from $\Psi_0$ by a ``soliton'' transformation.  See \cite{bz} for more details. A general
solution is obtained by the addition of any number of solitons and anti-solitons. An $n$-soliton transformation, however, can be realized
as the composition of $n$ 1-soliton transformations. To derive some of the results we will need later, it will be more convenient for
us to work with 1-soliton transformations. Repeated applications of 1-soliton transformations will allow us to recover the most general case.

The 1-soliton BZ transformation is parametrized by the  ``soliton position''
\be
\mu = \sqrt{r^2 + (z-p)^2}-(z-p)
\ee
where $p$ is a real number, and by the BZ parameters, contained in the constant vector $\bar m_a$ ($a=1,2,3$). The following results
also apply, with no modification, if one replaces $\mu$ with $\tilde\mu$
\be
\tilde\mu = -\sqrt{r^2 + (z-p)^2}-(z-p)
\ee
This corresponds to the addition of an ``anti-soliton'' rather than a soliton. Note that both $\mu$ and
$\tilde \mu$ satisfy the differential relations
\be
\partial_r \mu = {2 r \mu\over r^2+\mu^2}\,,\quad \partial_z \mu = -{2\mu^2\over r^2+\mu^2}
\label{mudiff}
\ee

Given these parameters, one defines the following $r$ and $z$ dependent vector $m_a$
\be
m_a = {\bar m}_b [\Psi^{-1}_0(\mu,r,z)]_{ba}
\ee
and the matrix $P$
\be
P_{ab}={[G_0]_{ac} m_c m_b\over [G_0]_{de} m_d m_e}
\ee  
$P$ has the property of a projector:
\be
P^2=P\,,\quad \mathrm{det} P = 0
\ee
One can now construct  the new solution $G$ as
\be
G=\Bigl(I-{r^2+\mu^2\over \mu^2}P\Bigr)G_0
\label{unphysical}
\ee
However, $G$ does not obey the required relation $\mathrm{det} G = -r^2$; instead one finds
\be
\mathrm{det} G = -{r^2\over \mu^2}\mathrm{det} G_0
\label{det}
\ee
There are several possible solutions to this problem. One is to re-scale the metric, or a sub-block of it, by a suitable factor. Another, which we will use in our application of Section \ref{sectfive}, is to act with a  series of BZ transformations, in which one subtracts and adds the same
soliton, but with different BZ parameters: since the relation (\ref{det}) does not contain the BZ parameters, such a series of transformations does not change the determinant of $G$. 

To compute the full metric one needs the factor $e^{2\nu}$, that solves Eq. (\ref{nueq}). One can derive an explicit identity that connects the factor $e^{2\nu}$ for the new metric $G$ to the factor $e^{2\nu_0}$ for
the seed $G_0$:
\be
e^{2\nu}= c_\nu e^{2\nu_0} {r \mu^2\over (r^2 +\mu^2)^2}[G_0]_{ab} m_a m_b
\ee
$c_\nu$ is a constant that can be fixed, for example, by requiring the right behavior at asymptotic infinity.

\subsection{Computing $\chi$ and $\kappa$}\label{sectfourtwo}
Given a metric $G$ produced by the IS construction, one would like to be able to apply an $SL(3,\mathbb{R})$ transformation to it. 
This requires solving the differential equations (\ref{dual}) and (\ref{kappadef}), in order to compute the potentials $V_a$ and
$\kappa$. In most practical cases, this is a daunting computation. In this subsection we provide a solution to this problem, by deriving a series of identities that capture the transformation rules of $V_a$ and $\kappa$ under a 1-soliton transformation. Using these
identities one can compute the $V_a$ and $\kappa$ associated to the metric $G$ from the corresponding objects for the seed
$G_0$.

Consider first a preliminary problem. The equations of motion (\ref{equv}) imply the existence
of a matrix $\Gamma$ such that
\be
\partial_z\Gamma=  -{1\over 2} U\,,\quad  \partial_r \Gamma ={1\over 2} V
\label{gammadef}
\ee
(the factor of $1/2$ is chosen for later convenience). 
 We will show later that $\Gamma$ is a useful quantity to compute and we would thus like to know how $\Gamma$ transforms under an IS transformation. Since $\Gamma$ is obtained by integrating $U$ and 
$V$, we need to work out their transformation laws first.  From the transformation law for $G$, Eq. (\ref{unphysical}), one finds
\bea
\!\!\!U \!\!\!&=&\!\!\! U_0 + 2 {\mu^2-r^2\over r^2+\mu^2} P - U_0 P - P U_0 + 2 P U_0 P -{\mu\over r} V_0 P +{r\over \mu} P V_0 +\Bigl({\mu\over r}-{r\over \mu}\Bigr) P V_0 P \nonumber\\
\!\!\!V \!\!\!&=&\!\!\! V_0 + 4 {r \mu\over r^2+\mu^2} P - V_0 P - P V_0 + 2 P V_0 P +{\mu\over r} U_0 P -{r\over \mu} P U_0 -\Bigl({\mu\over r}-{r\over \mu}\Bigr) P U_0 P\nonumber\\
\eea
Using the differential relations for $\mu$, given in Eq. (\ref{mudiff}), and the relations
\bea
\partial_r P\!\!\! &=&\!\!\! {1\over r} U_0 P-{r U_0 + \mu V_0\over r^2+\mu^2} P - P {r U_0 +\mu V_0\over r^2+\mu^2}+{r^2-\mu^2\over r(r^2+\mu^2)}PU_0 P + {2\mu\over r^2+\mu^2}PV_0 P\nonumber\\
 \partial_z P\!\!\! &=&\!\!\! {1\over r} V_0 P-{r V_0 - \mu U_0\over r^2+\mu^2} P - P {r V_0 -\mu U_0\over r^2+\mu^2}+{r^2-\mu^2\over r(r^2+\mu^2)}PV_0 P - {2\mu\over r^2+\mu^2}PU_0 P\nonumber\\
 \label{pdiff}
\eea
that follow from the definition of $P$, one can verify that $U$ and $V$ can be rewritten as
\bea
U&=& U_0 -\partial_z\Bigl({r^2+\mu^2\over \mu} P \Bigr)\nonumber\\
V&=& V_0 +\partial_r\Bigl({r^2+\mu^2\over \mu} P \Bigr)
\eea
This proves the following transformation rule for $\Gamma$
\be
\Gamma = \Gamma_0 +{r^2+\mu^2\over 2 \mu} P
\label{gammares}
\ee
where $\Gamma_0$ satisfies
\be
\partial_z\Gamma_0=  -{1\over 2} U_0\,,\quad  \partial_r \Gamma_0 ={1\over 2} V_0
\ee

We would now like to relate $\Gamma$ to the quantities we were originally looking for, i.e. $V_a$ and $\kappa$.
 It is convenient to change from the coordinates $(t,\phi,\psi)$ to $(t,\phi_+,\phi_-)$, where
\be
\phi_\pm = \psi\pm \phi
\ee
We will denote by $M_{IJ}$, with $I,J=0,+,-$, the components of any matrix $M$  in the $t,\phi_+,\phi_-$ base.

By comparing the equations satisfied by the potentials $V_0$ and $V_1$, with the equation defining 
$\Gamma$, (\ref{gammadef}), one finds that
\be
V_0 = {\Gamma_{0-}\over \ell}+c_0\,,\quad V_1 = {\Gamma_{+-}\over \ell^2}+c_1
\label{v0v5}
\ee
$c_0$ and $c_1$ are constants, that are fixed by the asymptotic boundary conditions. 
In a similar way, if we write the matrix $\kappa$ as
\be
\kappa = \begin{pmatrix}\kappa_{00} &\kappa_{01}&\kappa_{02}\cr
\kappa_{10}&\kappa_{11} & \kappa_{12}\cr
\kappa_{20}&\kappa_{21}&\kappa_{22} \end{pmatrix} d{\tilde \phi}
\ee
one finds that
\bea
\kappa_{00}&=&V_0 \omega^0 + {\Gamma_{00}\over \ell}+c_{00}\nonumber\\
\kappa_{01}&=&V_1 \omega^0 + {\Gamma_{+0}\over\ell^2}+c_{01}\nonumber\\
\kappa_{02}&=&- \omega^0 \nonumber\\
\kappa_{10}&=&V_0 \omega^1 + \Gamma_{0+}+c_{10}\nonumber\\
\kappa_{11}&=&V_1 \omega^1 + {\Gamma_{++}\over\ell}+c_{11}\nonumber\\
\kappa_{12}&=&- \omega^1 \nonumber\\
\kappa_{20}&=&V_0 (V_0 \omega^0 +V_1 \omega^1)+{1\over 2\ell^2} (\Gamma \sigma \Gamma)_{0-} + {{\tilde \kappa}_{0-}\over\ell^2}\nonumber\\
&&+{c_0\over \ell} (\Gamma_{00}-\Gamma_{--}-z)+c_1 \Gamma_{0+}+ c_{20} \nonumber\\
\kappa_{21}&=&V_1 (V_0 \omega^0 +V_1 \omega^1)+{1\over 2\ell^3} (\Gamma \sigma \Gamma)_{+-} + {{\tilde \kappa}_{+-}\over\ell^3}\nonumber\\
&&+{c_1\over \ell} (\Gamma_{++}-\Gamma_{--}-z)+{c_0\over \ell^2} \Gamma_{+0}+c_{21}\nonumber\\
\kappa_{22}&=& -V_0 \omega^0-V_1 \omega^1 -{\Gamma_{00}\over\ell}-{\Gamma_{++}\over\ell}-c_{00}-c_{11}
\label{kappamess}
\eea
where $\sigma$ is a matrix that is given, in the $t,\phi_+,\phi_-$ basis, by
\be
\sigma=\begin{pmatrix}1&0&0\cr 0&1&0\cr 0&0&-1\end{pmatrix} 
\ee
and $c_{ij}$ are further constants needed to satisfy the asymptotic boundary conditions.
$\tilde\kappa_{IJ}$ are the components of a new matrix, defined by the following differential relations
\bea
\partial_r {\tilde\kappa} &=& -{r\over 2} \partial_z \Gamma + \half\com{\partial_r \Gamma }{\Gamma} \nonumber\\
\partial_z {\tilde\kappa} &=& {r\over 2} \partial_r \Gamma -\Gamma + \half\com{\partial_z \Gamma }{\Gamma} \
\label{ktildeeq}
\eea
To see that this system of equations is integrable, note that the equation defining $\Gamma$, Eq. (\ref{gammadef}), implies
\be
\Bigl(\partial_r ^2 +\partial_z ^2-{1\over r}\Bigr) \Gamma - {2\over r} \com{\partial_r \Gamma}{\partial_z\Gamma } =0
\ee
This provides the integrability condition for the system (\ref{ktildeeq}). Not only is this system integrable, but one can also exhibit an explicit solution for it, in the case in which the metric is obtained by applying a 1-soliton transformation to a seed metric. Let $\tilde \kappa_0$ denote the solution of Eqs. (\ref{ktildeeq}) for the seed metric $G_0$. Then, using our previous result for $\Gamma$, given in Eq. (\ref{gammares}), and the differential relations for $\mu$ and $P$ given in (\ref{mudiff}) and (\ref{pdiff}), it is
not difficult to verify that 
\be
\tilde\kappa = \tilde\kappa_0- {(r^2+\mu^2)(r^2-\mu^2)\over 8 \mu^2} P +{r^2+\mu^2\over 4 \mu} \com{P}{\Gamma_0} 
\label{ktilderes}
\ee
solves the system (\ref{ktildeeq}).

Let us summarize our findings. Eqs. (\ref{v0v5}) and (\ref{kappamess}) show that both the potentials $V_0$ and $V_1$ and the matrix $\kappa$ are (algebraically) determined in terms of
the gauge fields $\omega^0$, $\omega^1$, the matrix $\Gamma$ and the matrix $\tilde \kappa$. Eqs. (\ref{gammares}) and (\ref{ktilderes}) provide explicit expressions for $\Gamma$ and $\tilde\kappa$ in terms of the analogous quantities for the seed solution and of the BZ parameters.  This completes our program of computing $\chi$ and $\kappa$ for any metric generated by the BZ technique, once the corresponding quantities for the seed metric are known.

\newsection{Static Black Ring in Taub-NUT Space}\label{sectfive}

In this section we apply the results of the previous section to present the construction of a static black ring in a KK-monopole background. The procedure can be summarized as follows: we first construct a modified seed metric which has the same rod structure as the static black ring except the orientation of the finite space-like rod is a generic mix of the two space-like directions, as in Figure 2(b). This solution is generated by applying the Belinski-Zakharov inverse scattering technique to a diagonal seed metric. In the resulting modified seed metric the finite space-like rod orientation is parametrized by one of the soliton parameters. Then, by applying the $D$ transformation to the modified seed solution we add KK-monopole charge and fix the remaining soliton parameter so that the final solution has the correct rod structure as depicted in Figure 2(a). 

The layout of this section is as follows. In Section \ref{sectfiveone} we construct the modified seed metric using the inverse scattering technique and the results of Section \ref{sectfour}. Then, in Section \ref{sectfivetwo} we apply the $D$ transformation to add the KK-monopole charge. We examine the rod structure of this solution and fix the remaining BZ parameter to produce the static black ring in Taub-NUT space. When possible we present explicit results though due to the complexity of the expressions many of the results have to be relegated to the appendices.
\begin{figure}[h]
\begin{center}
$
\begin{array}{c@{\hspace{0.2in}}c}
\includegraphics[totalheight=3 cm]{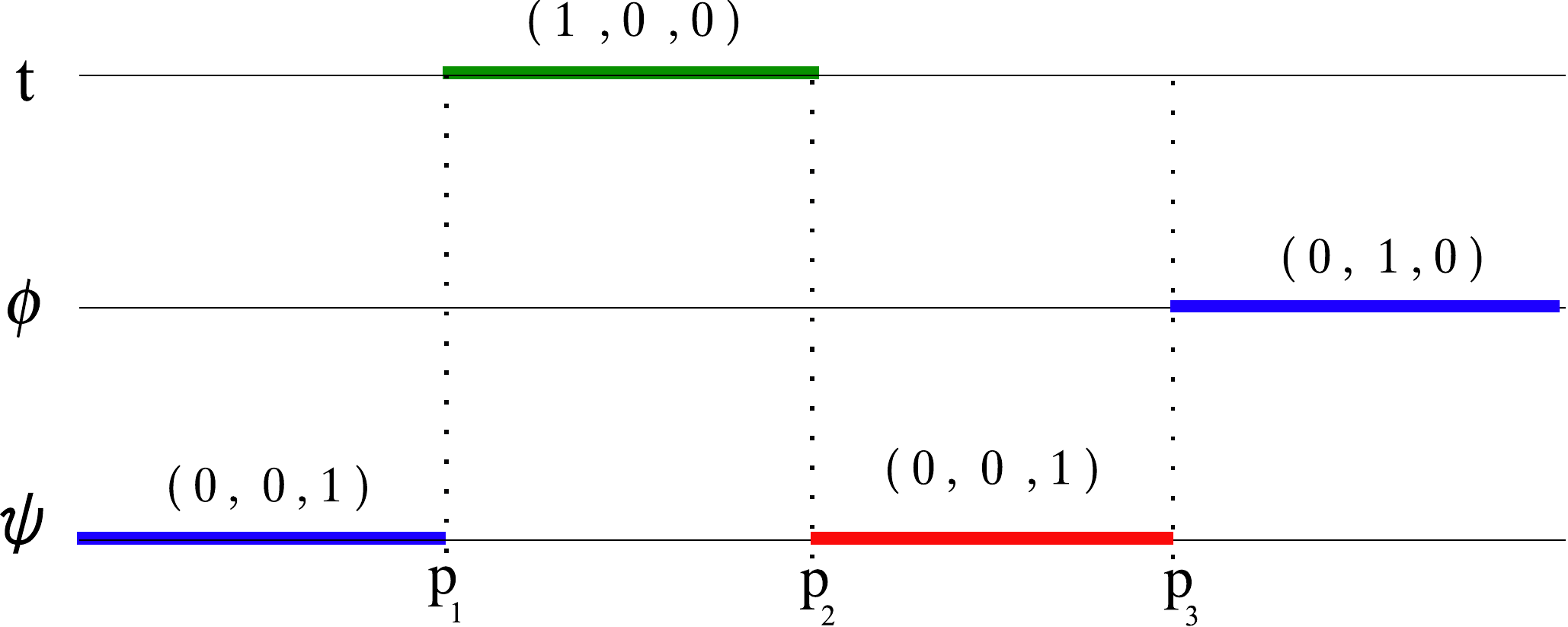}
&
\includegraphics[totalheight=3 cm]{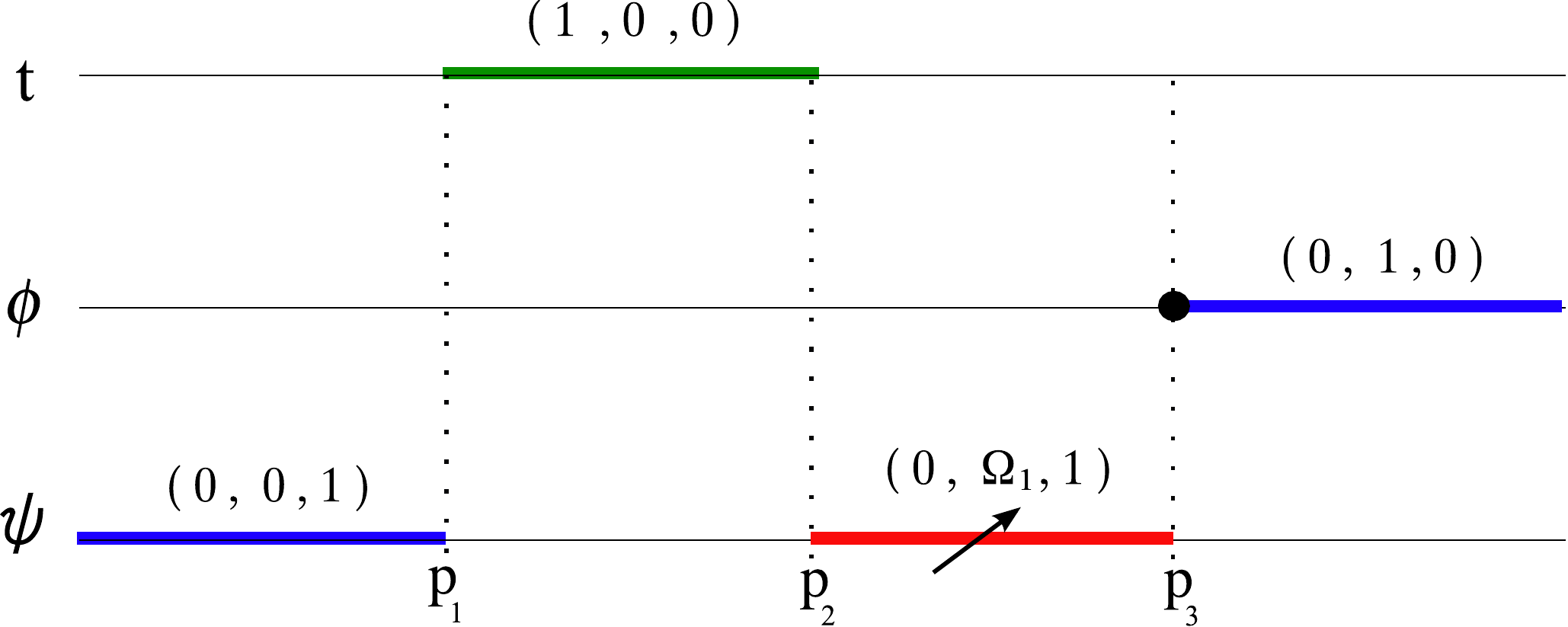}\\
\\
\mbox{(a)} & \mbox{(b)}
\end{array}
$
\end{center}
\caption{(a) The rod structure of the static black ring. (b) The rod structure of the modified seed metric in which the finite space-like rod is rotated. }
\end{figure}

\subsection{Generating the Modified Seed Metric: BZ Soliton Transformations}\label{sectfiveone}
In this section we present the construction of the modified seed solution in which the eigenvector of the finite space-like rod is generically oriented between $\phi$ and $\psi$ as shown in Figure 2(b).  To construct this metric we apply the method of Pomeransky~\cite{pomeransky} to a seed metric that coincides with the static black ring, with the rod structure depicted in Figure 2(a). The seed metric and  the corresponding $\Gamma$ and $\tilde{\kappa}$ are given by,
\bea
ds^2_0&=&G^{(0)}_{IJ}dx^Idx^J+e^{2\nu_0}(dr^2+dz^2)\\
G^{(0)}&=& \mbox{\rm{diag}} \{-\frac{\mu_1}{ \mu_2} ,  \mu_3 , -\frac{\mu_{2}\tilde{\mu}_{3}}{\mu_1} \}\\
e^{2\nu_0}&=& \frac{\mu_3 (r^2+\mu_1\mu_2)^2 (r^2+ \mu_2 \mu_3)}{(r^2+\mu_1^2)(r^2+\mu_2^2)(r^2+ \mu_3^2)(r^2+ \mu_1\mu_3)} \\
\Gamma^{(0)} &=& \frac{1}{2} \mbox{\rm{diag}} \{ \tilde{\mu}_{1} -\tilde{\mu}_{2}, \tilde{\mu}_{3}, \mu_3 +\tilde{\mu}_{2} -\tilde{\mu}_1 \} \\
\tilde{\kappa}^{(0)} &=& \frac{1}{8} \mbox{\rm{diag} } \{ \mut_1^2 -\mut_2^2 , \mut_3^2 , \mu_3^2 +\mut_2^2 -\mut_1^2 \}
\eea
where the elements of the above matrices are the $tt, \,\,\phi\phi,\,\, \psi\psi$ components and the indices $I$ and $J$ lie in $\{0,1,2\}$. The $\mu_i$'s and $\mut_i$'s  are defined as
\be
\mu_i=\sqrt{r^2+(z-p_i)^2}-(z-p_i ),\ \mut_i = -\sqrt{r^2+(z-p_i)^2}-(z-p_i )\,\,\,
\ee

Note that throughout the following analysis the time-like components will not mix with the space-like components. Specifically, during the soliton transformations the BZ parameters will be chosen so that only the $\phi$ and $\psi$ components are affected. As a result the metric can be decomposed into a $1\times1$ time-like block times a $2 \times 2$ space-like block. This decomposition will be denoted by,
\bea
G &=& \begin{pmatrix} G_{tt} &0&0\cr
0&G_{\phi\phi} & G_{\phi\psi}\cr
0&G_{\psi\phi}&G_{\psi\psi} \end{pmatrix}=\begin{pmatrix} G_{tt} & 0 \cr
0 & G_2  \end{pmatrix}
\eea
\bea
\Gamma &=& \begin{pmatrix} \Gamma_{tt} &0&0\cr
0&\Gamma_{\phi\phi} &\Gamma_{\phi\psi}\cr
0&\Gamma_{\psi\phi}&\Gamma_{\psi\psi} \end{pmatrix}=\begin{pmatrix} \Gamma_{tt} &0 \cr
0 &\Gamma_2  \end{pmatrix}
\eea
where $G_2$ and $\Gamma_2$ denote the $2 \times 2$ space-like blocks.

After applying the soliton transformations we will need to compute the $\Gamma$ and $\tilde{\kappa}$ matrices for the modified seed metric. As explained earlier, this task is easily accomplished by using the formalism developed in Section \ref{sectfourtwo}. 

The procedure used to generate the modified seed solution with a rotated finite space-like rod is summarized in the following steps:
\begin{itemize}
\item{Step 1:}  Remove an anti-soliton at $z=p_3$ with BZ parameters $m^{(1)}=(0,1,0)$. The resulting metric is
\be
\mbox{\rm{diag}} \{ -\frac{\mu_1}{\mu_2}, -\frac{r^2}{\mu_3}, \frac{r^2\mu_{2}}{\mu_1\mu_3} \}
\ee
\item{Step 2:} Rescale the metric by $\lambda= -\frac{\mu_3}{r^2}$ to get
\be
\mbox{\rm{diag}} \{ \frac{\mu_1 \mu_3}{r^2 \mu_2}, 1, -\frac{\mu_{2}}{\mu_1} \}
\ee

\item Step 3:  Put back the anti-soliton removed in the Step 1  at $z=p_3$ with BZ parameters $m^{(2)}=(0,1,a)$. Let the metric produced in this step be denoted  as $G^{\mathrm{unph}}$ and the corresponding $\Gamma$ matrix as $\Gamma^{\mathrm{unph}}$. 

\item Step 4:  Undo the rescaling performed in Step 2, i.e. multiply the metric by $\frac{1}{\lambda}$. The resulting metric and $\Gamma$ will be denoted by $G^{\mathrm{ph}}$ and $\Gamma^{\mathrm{ph}}$ respectively. 
The $\Gamma$ matrix transforms under this rescaling  as,
\bea
\Gamma^{\mathrm{ph}}&=& \Gamma^{\mathrm{unph}}+ \frac{\mu_3}{2}\mathbb{I}_3
\eea
where $\mathbb{I}_3$ is the $3 \times 3$ identity matrix. It can be shown that the relevant component of $\tilde{\kappa}$ is not affected by the rescaling transformation. Finally the rescaling changes the conformal factor as
\be
\nu_{\mathrm{ph}}=\nu_{\mathrm{unph}}+\Delta \nu
\ee
where $\Delta\nu$ solves the following differential relations
\bea
\partial_r \Delta\nu&=& -\frac{(\partial_r\log\lambda )\mbox{\rm{Tr}}U }{4}+\frac{(\partial_z\log\lambda ) \mbox{\rm{Tr}}V }{4}+\frac{r}{4}\left[ (\partial_r\log\lambda )^2-(\partial_z\log\lambda )^2\right]  \nonumber\\
\partial_z \Delta\nu&=& -\frac{(\partial_r\log\lambda ) \mbox{\rm{Tr}} V }{4}+\frac{(\partial_z\log\lambda ) \mbox{\rm{Tr}}U }{4}+\frac{r}{2}\left[ (\partial_r\log\lambda )(\partial_z\log\lambda )\right]  \nonumber
\eea
\end{itemize}
The final result of the above procedure, i.e. the modified seed metric with a rotated finite space-like rod, Figure 2(b), is given by: 
\be
ds^2=  -\frac{\mu_1}{\mu_2}dt^2+ 2G_{\phi\psi}d\phi d\psi +G_{\phi\phi}d\phi^2+G_{\psi\psi}d\psi^2+e^{2\nu}(dr^2+dz^2)
\label{modifiedseed}
\ee
\bea
G_{\phi\phi}&=& \frac{1}{\Delta_G}\left(\mu_1\mu_3^2\RR 2 3 ^2 +a^2 r^2 \mu_2 \RR 1 3 ^2  \right)\nonumber\\
G_{\psi\psi}&=& \frac{\mu_2}{\mu_1\Delta_G} \left( r^2\mu_1\RR 2 3 ^2 +a^2 \mu_2 \mu_3^2 \RR 1 3 ^2  \right)\nonumber\\
G_{\phi\psi}&=& \frac{a}{\Delta_G}  \mu_2  \RR  3 3 \RR 1 3 \RR 2 3    \nonumber\\
e^{2\nu}&=&\frac{\Delta_{G} \RR 1 2 ^2}{\mu_1 \RR 1 1 \RR 2 2 \RR 3 3 \RR 1 3 \RR 2 3} \nonumber
\eea
where
\be
\Delta_G = \mu_3( \mu_1 \RR 2 3 ^2  -a^2\mu_2 \RR 1 3 ^2)
\ee
and in order to simplify the expressions, we have introduced the following convenient notation \cite{evslin}:
\be
\mathcal{R}_{ij}  =  (r^2+\mu_i\mu_j)
\ee
The matrices $\Gamma$ and $\tilde\kappa$ associated to the metric above can be easily derived by using Eq. (\ref{gammares}) and Eq. (\ref{ktilderes}) respectively. From this data, we can then construct $\chi$ and $\kappa$ needed for the next step by using Eq. (\ref{v0v5}) and (\ref{kappamess}).

The BZ parameter $a$ determines the orientation of the eigenvectors associated with the space-like rods.
In the basis $(\partial_t,\partial_{\phi}, \partial_{\psi})$, the orientation of the left semi-infinite rod is
\be
(0,-a,1)
\ee
that of the right semi-infinite rod is
\be
(0,1,-a)
\ee
and that of the finite space-like rod is
\be
\left(0,-a {p_3-p_1\over p_3-p_2},1\right )
\ee
The fact that the semi-infinite rods do not have the standard orientation $(0,0,1)$ and $(0,1,0)$
simply means that a change of coordinates is needed to bring the metric into explicitly flat form at asymptotic infinity. The relevant point to note is that the left semi-infinite rod and the finite space-like rod are
not parallel for $a\not =0$, which is the basic property we require from the modified seed metric. For $a=0$ the two rods become parallel and the solution above reduces to the static black ring. 

\subsection{Adding KK-Charge: applying the $D$ transformation}\label{sectfivetwo}
Having obtained the metric with rod structure of Figure 2(b), we can now add KK-charge by the application of the $\slthree$ transformation $D$: 
\bea
\chi &\rightarrow& \bar\chi=D\chi D^{T}\nonumber\\
\kappa &\rightarrow& \bar\kappa=D\kappa D^{T}
\label{Dsec5}
\eea
We do not give the explicit expressions for $\bar\chi$ and $\bar\kappa$ as they can be easily computed from Eq. (\ref{Dsec5}) and the $\chi$ and $\kappa$ corresponding to the metric in Eq. (\ref{modifiedseed}).  The metric of the black ring with KK-charge can now be extracted from $\bar\chi$ and $\bar\kappa$. The metric can be written as
\be
ds^2_5=\bar\lambda_{00}dt^2+\bar\lambda_{11}(d\xi^1+\bar\omega^1)^2+\frac{1}{\bar\tau}ds_3^2
\ee
where
\bea
&&\bar\lambda_{00}=\bar\chi_{00} =-\frac{\mu_1}{\mu_2},\ \bar\lambda_{11}=\bar\chi_{11} -\frac{\bar{\chi}_{12}^2 }{\bar{\chi}_{22} },\ \bar\omega^1=-\bar{\kappa}_{12}\\
&& \tau = -\frac{1}{\chi_{22} },\ \bar{\tau}= -\frac{1}{\bar{\chi}_{22} },\ ds_3^2= \tau e^{2\nu}(dr^2+dz^2)+\frac{r^2}{4}d\tilde{\phi}^2
\eea
and
\be
\xi^1 = \ell(\psi+\phi),\ \tilde{\phi} =\psi-\phi
\ee
The explicit metric functions are given in Appendix \ref{appendixD}. 

The solution we have constructed above depends on the parameter $a$. The value of this parameter is
fixed by requiring that the eigenvector associated to the finite space-like rod $[p_2,p_3]$ be parallel to the eigenvector associated to the semi-infinite rod $(-\infty,p_1]$ (see Figure 2(a)). It is important to note that this in itself does \emph{not} guarantee that the horizon topology is $S^{2}\times S^{1}$. It is possible to have singularities at the rod junctions which cause the norm of the degenerating cycle to jump at the intersection.\footnote{We thank Joan Camps, Roberto Emparan and Pau Figueras for pointing out the existence of this pathology to us.} We will check later that this pathology is absent for the geometries we construct.

 The eigenvectors associated to each rod are easily computed using the relation (\ref{spaceeigen}). Let us denote by $v_<$, $v$ and
$v_>$ the eigenvectors associated to the rods $(-\infty,p_1]$, $[p_2,p_3]$ and $[p_3,\infty)$. Let us also use the following convenient parametrization for the points $p_i$:
\be
p_1= - c k^2\,,\quad p_2 = c k^2\,,\quad p_3 = k^2
\ee
where $c$ is a dimensionless parameter with $0<c<1$, and $k$ sets the length scale of the solution. The Taub-NUT geometry is characterized by a second length scale, $\ell$, and it will be useful to introduce the dimensionless ratio
\be
\khat={k\over \ell}
\ee
The rod eigenvectors are
\be
v_< = (0,Q,1)\,,\quad v = (0,\tilde Q,1)\,,\quad v_>=(0,-Q,1)
\ee
where $Q$ is given by 
\be
Q= \ell\, {4(1-a)+c (c-2 -a (c+2))\khat^4\over 8 (1+a)}
\ee
and $\tilde{Q}$ by
\be
\tilde Q=Q - \ell \frac{c( (1-c) \khat^2 - a (2-\khat^2) ) (2+\khat^2 + a \khat^2 (1+c))}{2 (1+a) (1-c+ a(1+c) )}
\ee
The condition which guarantees that $v_{<}$ is parallel to $v$ is thus $Q=\tilde Q$. This condition gives two solutions for $a$:
\bea
a_{0} = \frac{(1-c)\khat^2}{2-\khat^2},\ a_1 = -\frac{2+\khat^2}{(1+c)\khat^2}
\label{avalue}
\eea
Let us consider first the solution $a_{0}$. It has the property that when $c\rightarrow 1$, $a_{0}$ vanishes and one recovers the known~\cite{rasheed} geometry of the five dimensional Schwarzschild black hole with KK-monopole charge, as expected. Furthermore, when $\ell \rightarrow \infty$ keeping $k$ and $c$ fixed, (i.e. $\khat \rightarrow 0$), $a_{0}$ vanishes and one recovers the static black ring in five dimensional flat space, as can be seen by using the general result of Appendix A. This limit corresponds to taking the radius of the KK-monopole to be much larger than the  mass scale of the underlying ring. Thus $a_{0}$ represents the class of solutions which continuously connects to the static black ring in flat space. The branch corresponding to $a_{1}$ is rather unphysical, in that the area of the horizon identically vanishes on this branch. Indeed the horizon area can be computed to be proportional to
\be
\frac{c^2 \khat^3 \ell^3(2 + \khat^2 +  a (1+c)\khat^2)}{\sqrt{1 + c}}
\ee
where the constant of proportionality depends on the periodicities chosen for $\tilde{\phi}$ and $\xi^1$. It is easy to see that for $a=a_1$ the area vanishes. In the following we discard this solution.

\subsubsection{Asymptotic limit and KK-monopole charge}\label{sectfivetwoone}
To study the asymptotic behaviour of the solution let us change to the coordinates $\rho$ and $\theta$ defined in Eq. (\ref{asympcoord}) and take the limit
$\rho\to\infty$. The metric becomes
\be
ds^2 = -dt^2 + (d\xi^1 + Q \cos2\theta d\tilde{\phi} )^2 + \frac{(1+a)^2\rho^2}{4\ell^2} (d\rho^2 + \rho^2 d\theta^2) + \frac{\rho^4}{16\ell^2} \sin^22\theta d\tilde{\phi}^2
\ee
The following change of coordinates
\be
R = \frac{(1+a) \rho^2}{4 \ell},\ \bar{\theta} = 2\theta,\ \bar{\phi} = \frac{\tilde{\phi}}{1+a}
\ee
brings the above metric to the standard Taub-NUT asymptotic form with KK-monopole charge
\be
Q_{KK} = (1+a) Q = \ell \frac{c^3 \khat^6 - 2c(2-c) \khat^4 - 4(2-c)\khat^2 +8 }{8(2-\khat^2)}
\ee 
It is important to note that the smallest zero of the above expression is at
\be
\khat^2 = \khat^2_{*} \equiv 2 \left( \frac{1-\sqrt{1-c^2}}{c^2} \right) < 2 \le \frac{2}{c} \label{kstar}
\ee 
Thus as $\khat$ varies from $0$ to $\khat_{*}$, $c$ varies between $0$ and $1$, and $\ell$ varies from $0$ to $\infty$, one covers the entire physical range of the parameters $k$, $c$ and $Q_{KK}$. Hence, the apparent singularity at $\khat=\sqrt{2}$ is not physical and is an artifact of the particular parametrization. 
\subsubsection{Geometry of the horizon}\label{sectfivetwotwo}

In the canonical coordinates for our geometry, the horizon corresponds to the region $r=0$ with $z \in [-c k^2,ck^2]$. It is convenient to define the following change of coordinates 
\be
z =c k^2 \cos\theta, \ \theta \in [0,\pi]
\ee
\be 
\xi^{1} = \ell \left[ \frac{(2 +c\khat^2 )(4-4\khat^2 + c^2 \khat^4)}{8(2-c \khat^2)} \hat{\phi} + \hat{\psi} \right], \ \tilde{\phi} = \hat{\phi}
\ee
The geometry of a spatial cross-section of the horizon in coordinates $\theta, \hat{\phi}$ and $\hat{\psi}$ is
\be
ds^2_H = \frac{c^2 \khat^2 (4- c^2 \khat^4)^2 }{4(1+c) (2-\khat^2)^2} \left[ d\theta^2 + g_{\hat{\phi}\hat{\phi} } (d\hat{\phi} + \mathcal{A} d\hat{\psi} )^2 + g_{\hat{\psi}\hat{\psi} } d\hat{\psi}^2 \right] 
\ee
with
\bea
g_{\hat{\phi}\hat{\phi} } &=& \frac{(1+c)(2-\khat^2)^2 (4 - 4 c \khat^2 \cos\theta + c^2 \khat^4)}{(4 - c^2 \khat^4)^2 (1-c \cos\theta)} \sin^2\theta\\
g_{\hat{\psi}\hat{\psi} } &=& \frac{16 (1+c) (2-\khat^2)^2 (1-c\cos\theta)}{c^2 (4-c^2 \khat^4)^2 (4- 4 c\khat^2 \cos\theta + c^2 \khat^4)}\\
\mathcal{A} &=& \frac{4(4 - 4\khat^2 + c^2 \khat^4)}{(4-c^2 \khat^4) (4 - 4 c\khat^2 \cos\theta + c^2 \khat^4)}
\eea
Let us examine the limits $\theta\rightarrow 0,\pi$. As $\theta \rightarrow 0$ we find 
\bea
g_{\hat{\phi}\hat{\phi} } &=& \frac{(1+c)(2-\khat^2)^2 }{(1-c)(2+ c\khat^2)^2} \theta^2 + \mathcal{O}(\theta^4)\\
g_{\hat{\psi}\hat{\psi} } &=& \frac{16 (1-c^2) (2-\khat^2)^2 }{c^2 (4-c^2 \khat^4)^2 (2-  c\khat^2)^2 } + \mathcal{O}(\theta^2)\\
\mathcal{A} &=& \frac{4(4 - 4\khat^2 + c^2 \khat^4)}{(4-c^2 \khat^4) (2 - 2 c\khat^2)^2} + \mathcal{O}(\theta^2)
\eea
and as $\theta \rightarrow \pi$ we find
\bea
g_{\hat{\phi}\hat{\phi} } &=& \frac{(2-\khat^2)^2 }{(2-c \khat^2)^2}  (\pi-\theta)^2 + \mathcal{O}((\pi-\theta)^4)\\
g_{\hat{\psi}\hat{\psi} } &=& \frac{16 (1+c)^2 (2-\khat^2)^2 }{c^2 (4-c^2 \khat^4)^2 (2+  c\khat^2)^2 } + \mathcal{O}((\pi-\theta)^2)\\
\mathcal{A} &=& \frac{4(4 - 4\khat^2 + c^2 \khat^4)}{(4-c^2 \khat^4) (2 + 2 c\khat^2)^2} + \mathcal{O}((\pi-\theta)^2)
\eea
It is clear from these limiting behaviours that the norm of the vector $\partial_{\hat{\phi}}$ vanishes at both $\theta=0$ and $\theta=\pi$. It can also be checked that the norm of $\partial_{\hat{\psi}}$ does not vanish for any $\theta \in [0,\pi]$. This shows that the topology of the horizon is $S^2 \times S^1$ as expected. 

In general however, the two sphere parametrized by $\theta$ and $\hat{\phi}$ is singular because the periodicity of $\hat{\phi}$ cannot be chosen so as to cancel the conical defects at both the north and the south poles of the $S^2$. This is similar to the case of a static black ring in flat space. However, it is possible to cancel the conical defect at both the poles, if one imposes a specific relation between $c$ and $\khat^2$. From the expressions given above, it is easy to see that this happens when
\be
\frac{1+c}{1-c} = \left( \frac{2+c\khat^2}{2-c\khat^2} \right)^2
\ee
The only allowed solution to this equation is $\khat=\khat_{*}$. However, as follows from the definition of $\khat_*$ in Eq. (\ref{kstar}), at this point $Q_{KK}$ vanishes. By considering the complete geometry, it can be shown that this solution is nothing but the four dimensional Schwarzschild black hole times a trivial $S^1$. The same conclusion can be reached by looking at the periodicities imposed on the coordinates by the degenerations happening at the space-like rods.

\newsection{Conclusions}
In this paper we have presented a method that allows one to add KK-monopole charge to any asymptotically flat five-dimensional stationary axisymmetric solution of pure gravity. We have applied the method to generate the solution representing the vacuum static black-ring in  the Taub-NUT geometry.

A natural extension of our work would be to construct the most generally rotating ring in Taub-NUT. Using the techniques presented in this paper this extension is in principle straightforward, though
computationally quite laborious. The first step would be to obtain the five-dimensional seed solution, the analogue of the solution depicted in Figure 2(b).  Subsequently, one can act on this seed with the
$\slthree$ transformation $D$.  The seed solution should, this time, represent a rotating geometry. One expects that the action of $D$ on such a  geometry will give rise to a NUT charge, which will have to be cancelled \cite{rasheed} by the action of a further $SO(2,1)$ transformation; let us call it $N$.  We have shown in Section \ref{sectthreetwo} that $D$ rotates the eigenvector associated to a finite space-like rod in the $\phi,\psi$ plane. In a similar way,  $N$ further rotates this eigenvector, introducing  a mixing with the time direction. Thus,
 in this more general case, both the finite time-like
and space-like rods of the seed solution should be allowed to have generic orientations in the $(t,\phi,\psi)$ space. The orientation of the finite time-like rod accounts for the two angular momenta of
the five-dimensional ring, of which one combination will become the KK electric charge after the action of  $D$. The generic orientation of the finite space-like rod is needed to counterbalance the rotation induced by the action of $D$ and $N$. In order to generate a seed solution
with these properties one could apply again the Inverse Scattering technique. The construction of the seed solution should be a simple generalization of the one utilized in \cite{pomeransky} to find the five-dimensional black ring with two angular momenta. Using the results of Section \ref{sectfourtwo}, the $\chi$ and $\kappa$ matrices needed for applying $D$ and $N$ can be computed from the data of the IS construction.

Another interesting generalization of our work would be to find the extension of the transformation $D$
to supergravity theories in five dimensions. Such a generalization is bound to exist: Indeed the
$\slthree$ group is contained in the U-duality group of five-dimensional supergravity theories compactified down to three dimensions \cite{hull, sen}. Thus we expect that the generalized $D$ should be found as an element of this duality group. An application of this symmetry group as a solution generating technique in minimal supergravity has recently been discussed in~\cite{clement}.
\section*{Acknowledgements}


The authors would like to thank Joan Camps, Roberto Emparan and Pau Figueras for very useful correspondence and discussions. We acknowledge support from the National Science and Engineering Research Council of Canada (NSERC) and the Canadian Institute for Advanced Research (CIAR). J.F was supported by an Ontario Graduate Scholarship. 

\section*{Appendices}
\begin{appendix}
\renewcommand{\theequation}{\Alph{section}.\arabic{equation}}
\setcounter{equation}{0}
\renewcommand{\thesubsection}{\Alph{section}.\arabic{subsection}}
\setcounter{subsection}{0}
\section{Recovering the 5D solution}\label{appendixA}
We have seen in Section~\ref{secttwo} that to any stationary axisymmetric solution, with five-dimensional flat boundary
conditions, one can associate, via the transformation $D$, a solution whose asymptotic boundary 
is four-dimensional flat space times a circle and which carries KK-monopole charge; this solution depends on an extra length scale, $\ell$, that
parametrizes the length of the KK circle and the KK-monopole charge. In this appendix we will elucidate the relation between the
five-dimensional and four-dimensional geometries, and show that the five-dimensional geometry
sits at the tip of the KK-monopole geometry. More precisely, consider the limit of the four-dimensional geometry in which $\ell$ is sent to infinity and the  coordinates $r$ and $z$ and all the other 
parameters of the solution are kept fixed: in this limit one recovers the five-dimensional solution. 

To take this limit one has to make explicit the $\ell$ dependence of the matrices $\chi$ and
$\kappa$. Since, in the five-dimensional solution, $\ell$ only appears via the coordinate $\xi^1=\ell(\psi-\phi)$, it is easy to see that $\chi$ and $\kappa$ have the following form
\bea
\chi&=&\begin{pmatrix}\tilde\chi_{00}&{1\over \ell} {\tilde\chi}_{01}&\ell \tilde\chi_{02}\cr
{1\over \ell} {\tilde\chi}_{01}&{1\over\ell^2}{\tilde\chi}_{11}&\tilde\chi_{12}\cr
\ell \tilde\chi_{02}&\tilde\chi_{12}&\ell^2 \tilde\chi_{22} 
\end{pmatrix}=\begin{pmatrix}\tilde\lambda_{00}-{\tilde V_0^2\over \tilde\tau^2}&{1\over \ell}
\Bigl(\tilde\lambda_{01} - {\tilde V_0\tilde V_1\over \tilde\tau}\Bigr) &\ell {\tilde V_0\over\tilde\tau}\cr
{1\over \ell} \Bigl(\tilde\lambda_{01} - {\tilde V_0\tilde V_1\over \tilde\tau}\Bigr) &{1\over\ell^2}\Bigl(\tilde\lambda_{11} - {\tilde V_1^2\over \tilde\tau}\Bigr) & {\tilde V_1\over\tilde\tau}\cr
\ell {\tilde V_0\over\tilde\tau}&{\tilde V_1\over\tilde\tau}&-{\ell^2 \over \tilde\tau}
\end{pmatrix} \nonumber \\
\kappa&=&\begin{pmatrix}{1\over\ell}\tilde\kappa_{00}&{1\over \ell^2} \tilde\kappa_{01}&
\tilde\kappa_{02}\cr \tilde\kappa_{10} & {1\over\ell} \tilde\kappa_{11}&\ell \tilde\kappa_{12}\cr
{1\over\ell^2}\tilde\kappa_{20}&{1\over\ell^3}{\tilde\kappa}_{21}&{1\over\ell}\tilde\kappa_{22}
\end{pmatrix} 
\eea 
Quantities carrying a tilde are functions of $r$, $z$ and other parameters of the solution, but not of $\ell$;
so our limit consists of taking $\ell$ large keeping tilded quantities fixed. The four-dimensional solution is described by the matrices $\bar\chi=D\chi D^T$ and $\bar\kappa=D \kappa D^T$: in this solution $\ell$ is a physical parameter, it appears non-trivially in the metric and cannot be reabsorbed by a change of coordinates. Explicitly one finds
\bea
&&\!\!\!\!\!\bar{\chi}_{00}= \tilde\chi_{00}\,,\quad 
\bar{\chi}_{01}={\ell\over \sqrt{2}} \Bigl({\tilde\chi}_{02} + {{\tilde\chi}_{01}\over \ell^2}\Bigr)\,,\quad 
\bar{\chi}_{02}={\ell\over \sqrt{2}}\Bigl({\tilde\chi}_{02} -{{\tilde\chi}_{01}\over \ell^2}\Bigr)\nonumber \\
&&\!\!\!\!\!\bar{\chi}_{11}={\ell^2\over 2}\Bigl({\tilde\chi}_{22}+2{{\tilde\chi}_{12}\over \ell^2}
+{{\tilde\chi}_{11}\over \ell^4}\Bigr)\,,\quad \bar{\chi}_{12}= {\ell^2\over 2}\Bigl({\tilde\chi}_{22}-
{{\tilde\chi}_{11}\over \ell^4}\Bigr)\,,\quad 
\bar{\chi}_{22}={\ell^2\over 2}\Bigl({\tilde\chi}_{22}-{2 {\tilde\chi}_{12}\over \ell^2}+{{\tilde\chi}_{11}\over \ell^4}
\Bigr)\nonumber\\
&&\!\!\!\!\!\bar{\kappa}_{02}={1\over \sqrt{2}} \Bigl(\tilde\kappa_{02} -{\tilde\kappa_{01}\over \ell^2}\Bigr)\,,
\quad\bar{\kappa}_{12} ={\ell\over 2}\Bigl(\tilde\kappa_{12}+{\tilde\kappa_{22}-\tilde\kappa_{11}\over \ell^2}-{\tilde\kappa_{21}\over \ell^4}\Bigr) 
\label{ell4D}
\eea
However in the large $\ell$ limit things simplify again. One can compute, from the expressions (\ref{ell4D}), the coefficients of the four-dimensional metric, and take the limit $\ell\to\infty$;  one obtains
\bea
\bar{\lambda}_{00}&=&\bar{\chi}_{00}-{(\bar{\chi}_{02})^2\over \bar{\chi}_{22}}={\lambda}_{00}+O\Bigl({1\over \ell^2}\Bigr)\nonumber\\
\bar{\lambda}_{01}&=&\bar{\chi}_{01}-{\bar{\chi}_{02}\bar{\chi}_{12}\over \bar{\chi}_{22}}=
\sqrt{2} {\lambda}_{01}+O\Bigl({1\over \ell^3}\Bigr)\nonumber\\
\bar{\lambda}_{11}&=&\bar{\chi}_{11}-{(\bar{\chi}_{12})^2\over \bar{\chi}_{22}}=
2 {\lambda}_{11}+O\Bigl({1\over \ell^4}\Bigr)\nonumber\\
\bar{\omega}^0&=&-\bar{\kappa}_{02}= {\omega^0\over \sqrt{2}} +O\Bigl({1\over \ell^2}\Bigr) \,,\quad 
\bar{\omega}^1=-\bar{\kappa}_{12}= {\omega^1\over 2} +O\Bigl({1\over \ell}\Bigr)
\eea
The relations above show that the large $\ell$ limit of the four-dimensional metric coincides with the 
original five-dimensional metric, after the change of coordinates
\be
\sqrt{2}(\psi+\phi) =\bar\psi+\bar\phi\,,\quad {\psi-\phi\over \sqrt{2}}= \bar\psi-\bar\phi 
\ee

To summarize, we have shown that for any stationary axisymmetric vacuum solution which
asymptotes to five-dimensional Minkowski space one can generate, via the $SL(3,\mathbb{R})$ transformation $D$, a ``four-dimensional'' solution, having $\mathbb{R}^{3,1}\times S^1$ as its asymptotic limit and carrying KK-monopole charge. In the limit in which the size of the KK-monopole is taken to infinity, and all other quantities are kept fixed,  the four-dimensional solution goes over to 
the original five-dimensional solution. This shows that the transformation $D$ generates a 1-parameter family of solutions, interpolating between five-dimensional and four-dimensional geometries.

\section{The value of $\kappa$ at a space-like rod}\label{appendixB}\setcounter{equation}{0}
We give here the values of constants $k_{ij}$ that appear in Eq. (\ref{kapparod}). We consider all the possible cases for space-like rods.

\begin{itemize}
\item $z\in(-\infty, p_<)$: 
\bea
&&k_{12}=1,\quad k_{22}=k_{11}=-{1\over 4}\sum_{i_0}L_{i_0}\nonumber\\
&&k_{21}={1\over 8}[p_>^2+p_<^2-\sum_{i_1}(p_{i_1+1}^2-p_{i_1}^2)+\sum_{i_2}(p_{i_2+1}^2-p_{i_2}^2)]- c^2
\eea
(the constant $c$ has been defined in (\ref{c})).

\item $z\in(p_>,+\infty)$:
\bea
&&\!\!k_{12}=-1,\quad k_{22}=k_{11}={1\over 4}\sum_{i_0}L_{i_0}\nonumber\\
&&\!\!k_{21}=-{1\over 8}[p_>^2+p_<^2-\sum_{i_1}(p_{i_1+1}^2-p_{i_1}^2)+\sum_{i_2}(p_{i_2+1}^2-p_{i_2}^2)]+ c^2
\eea
\item $z\in(p_{i_1},p_{i_1+1})$:
\bea
&&\!\!k_{12}=-1\\
&&\!\!k_{22}=-{1\over 4}[3p_>+p_<- 4 p_{i_1}-3 L_{i_1}+\sum_{j_1>i_1}L_{j_1}-3 \sum_{j_1>i_1+1}L_{j_1}+\sum_{j_2}L_{j_2}]\nonumber\\
&&\!\!k_{11}={1\over 4}[p_>-p_<-\sum_{j_1} L_{j_1} -  \sum_{j_2<i_1}L_{j_2} + 3 \sum_{j_2>i_1+1}L_{j_2}]\nonumber
\eea
\bea
&&\!\!k_{21}={1\over 8}[p_>^2-p_<^2-p_{i_1}^2-p_{i_1+1}^2+
\sum_{j_1<i_1}(p_{j_1+1}^2-p_{j_1}^2) - \sum_{j_1>i_1+1}(p_{j_1+1}^2-p_{j_1}^2)\nonumber\\
&&\qquad\qquad -\sum_{j_2<i_1}(p_{j_2+1}^2-p_{j_2}^2) + \sum_{j_2>i_1+1}(p_{j_2+1}^2-p_{j_2}^2)]\nonumber\\
&&\!\!\qquad+{1\over 16}[p_>-p_<-p_{i_1}-p_{i_1+1} + \sum_{j_1<i_1}L_{j_1} - \sum_{j_1>i_1+1}L_{j_1}- \sum_{j_2<i_1}L_{j_2} + \sum_{j_2>i_1+1}L_{j_2} ]^2\nonumber
\eea

\item $z\in(p_{i_2},p_{i_2+1})$:
\bea
&&\!\!k_{12}=1\\
&&\!\!k_{22}=-{1\over 4}[p_>+3 p_<- 4 p_{i_2}- L_{i_2}-\sum_{j_1}L_{j_1}+3 \sum_{j_2<i_2}L_{j_2}-\sum_{j_2>i_2+1}L_{j_2}]\nonumber\\
&&\!\!k_{11}=-{1\over 4}[p_>-p_<+3\sum_{j_1<i_2} L_{j_1} -  \sum_{j_1>i_2+1}L_{j_2} -  \sum_{j_2}L_{j_2}]\nonumber\\
&&\!\!k_{21}={1\over 8}[p_>^2-p_<^2+p_{i_2}^2+p_{i_2+1}^2+
\sum_{j_1<i_2}(p_{j_1+1}^2-p_{j_1}^2) - \sum_{j_1>i_2+1}(p_{j_1+1}^2-p_{j_1}^2)\nonumber\\
&&\qquad\qquad -\sum_{j_2<i_2}(p_{j_2+1}^2-p_{j_2}^2) + \sum_{j_2>i_2+1}(p_{j_2+1}^2-p_{j_2}^2)]\nonumber\\
&&\!\!\qquad-{1\over 16}[p_>-p_<+p_{i_2}+p_{i_2+1} + \sum_{j_1<i_2}L_{j_1} - \sum_{j_1>i_2+1}L_{j_1}- \sum_{j_2<i_2}L_{j_2} + \sum_{j_2>i_2+1}L_{j_2} ]^2\nonumber
\eea

\end{itemize}

 \section{Static Black Ring in Taub-NUT Metric Functions}\setcounter{equation}{0}
\label{appendixD}
In this Appendix we give the metric representing the static black ring in Taub-NUT space as generated by the procedure described in Section \ref{sectfive}. Note that throughout this appendix the length scale $\l$ has been set to unity in order to simplify the expressions. The length scale can be restored by dimensional analysis. Also, we will use the following notation \cite{evslin}, 
\bea
\mathcal{R}_{ij} & = & (r^2+\mu_i\mu_j)\nonumber\\
\bar{ \mathcal{R}}_{ij} &= & (r^2-\mu_i\mu_j) \nonumber\\
\mathcal{D}_{ij} & = & (\mu_i-\mu_j)\nonumber\\
\Delta p & = & p_1-p_2 \nonumber
\eea

The metric of static black ring in TN space can be written as:
\be
ds^2=\lambda_{00}dt^2+\bar\lambda_{11}(d\xi^1+\bar\omega^1)^2 +\frac{1}{\bar\tau}\left[ \tau e^{2\nu}(dr^2+dz^2) +\frac{r^2}{4}d\tilde\phi^2  \right] 
\ee
where
\bea
\lambda_{00}&=&-\frac{\mu_1}{\mu_2}\\
\bar\tau&=&\lambda_{00}\bar\lambda_{11}\\
\bar\lambda_{11}&=&\frac{   N_{\lambda}^{(0)}+N_{\lambda}^{(1)}a+N_{\lambda}^{(2)}a^2         }{M_{\lambda}^{(0)}+M_{\lambda}^{(1)}a+M_{\lambda}^{(2)}a^2}\\
\bar\omega^1&=&\frac{N_{\omega}^{(0)}+N_{\omega}^{(1)}a+N_{\omega}^{(2)}a^2+N_{\omega}^{(3)}a^3}{(1+a)(M_{\omega}^{(0)}+M_{\omega}^{(1)}a+M_{\omega}^{(2)}a^2)}d\tilde\phi\\
e^{2\nu}&=&\frac{\mu_3\RR 1 2^2(\mu_1\RR 2 3 ^2-a^2\mu_2\RR 1 3 ^2)}{\mu_1\RR 1 1 \RR 1 3 \RR 2 2\RR 2 3 \RR 3 3}\\
\tau&=& \frac{\mu_1\RR 2 3 ^2(\mu_1\mu_3^2+\mu_2r^2) +2a\mu_1\mu_2\RR 1 3 \RR 2 3 \RR 3 3+a^2\mu_2\RR 1 3 ^2(\mu_2\mu_3^2+\mu_1r^2) }{4\mu_1\mu_2\mu_3\RR 2 3^2-4a^2\mu_2^2\mu_3\RR 1 3^2}
\eea
The coefficients in the above expressions for $\bar\lambda_{11}$ and $\bar\omega^1$ are,
\bea
N_{\lambda}^{(0)}&=& 8\mu_1^2 \mu_2^2\mu_3 \RR 2 3 ^2 (\mu_2r^2 +\mu_1\mu_3^2) \\
\nonumber\\
N_{\lambda}^{(1)}&=&16\mu_1^2\mu_2^3\mu_3\RR 1 3 \RR 2 3 \RR 3 3\\
\nonumber\\
N_{\lambda}^{(2)}&=& 8\mu_1\mu_2^3\mu_3\RR 1 3 ^2 (\mu_1 r^2 +\mu_2\mu_3^2)\\
\nonumber\\
M^{(0)}_{\lambda}&=&\mu_1^2\RR 2 3 ^2\left[ \mu_1\mu_2^2\mu_3^2 \pp^2 +16\mu_1\mu_2^2\mu_3^2+8\mu_1\mu_2^2\mu_3\RR 3 3 +4\mu_2\mu_3^2\DD 2 1 \tRR 1 2  \right.  \nonumber \\ &&    \left. \quad   -\mu_1\mu_2^2\mu_3\DD 2 1 \RR 3 3   +\mu_2\mu_3^3\DD 2 1 \RR 1 3 -\mu_2r^4\DD 2 1 \DD 2 3 - \mu_2\mu_3^2 r^2\DD 2 1 ^2     \right]\\
\nonumber\\
M^{(1)}_{\lambda}&=& \mu_1\mu_2\RR 1 3 \RR 2 3 \RR 33 \left[ 8\mu_1^2\mu_2\mu_3 + 8\mu_1\mu_2^2\mu_3 -\mu_1\DD 2 1 \DD 2 3 \RR 2 3\nonumber \right. \\ &&\quad \quad\quad\quad\quad\quad\quad  \left. -\mu_2 \DD 2 1\DD 1 3 \RR 1 3    \right]\\
\nonumber\\
M^{(2)}_{\lambda}&=& \mu_1\mu_2^2\RR 1 3^2\left[   -\pp^2\mu_1\mu_2\mu_3^2  -16\mu_1\mu_2\mu_3^2-4\mu_2\mu_3^2\RR 1 1   +8\mu_1\mu_2\mu_3\RR 3 3   \right. \nonumber\\ &&\left.  \quad\quad\quad\,  +  4\mu_1\mu_3^2\RR 2 2+\DD 1 2\DD 1 3 \RR 2 3 \RR 3 3       \right]\\
\nonumber\\
N^{(0)}_{\omega}&=& \mu_1\RR 2 3 ^2 \left[   16\mu_1\mu_2^2\mu_3 r^2 -16\mu_1^2\mu_2\mu_3^3 +16\mu_1\mu_2\mu_3^2 r^2\DD 2 1  +      8\mu_1\mu_3^3r^2\DD 1 2     \right. \nonumber\\ &&\left.  +8\mu_2\mu_3 r^4 \DD 2 1            +2r^2\DD 2 1 \DD 2 3 \RR 3 3 \RR 1 3       +\mu_2\pp (16\mu_1\mu_2\mu_3 r^2 -2\mu_1\mu_3 r^2 \RR 3 3  \right. \nonumber\\ &&\left.+2\mu_1\mu_2r^2\RR 2 3   +2\mu_1^2\mu_3^3\DD 2 3  +2\mu_1r^2\RR 1 3 \DD 2 3         -2\mu_2r^2\DD 1 3\RR 1 3           )   \right. \nonumber\\ &&\left.   +\mu_1\mu_2\mu_3\pp^2 (\mu_1\mu_3^2+3\mu_2r^2)    \right] 
\eea
\bea
N^{(1)}_{\omega}&=&\RR 2 3 \left[      16\mu_1^2\mu_2\mu_3\RR 2 3 (\mu_2 r^2-\mu_1\mu_3^2)   +16\mu_1^2\mu_2\mu_3^2r^2\DD 2 1 \RR 2 3 -8\mu_1^2\mu_3^3r^2\DD 2 1\RR 2 3    \right. \nonumber\\ &&\left.          + 8\mu_1\mu_2\mu_3r^4\DD 2 1 \RR 1 3 +4\mu_1r^2\DD 2 1 \DD 2 3 \RR 1 3 \RR 2 3 \RR 3 3 +2\mu_2r^2 \DD 2 1 \DD 1 3 \RR 1 3 ^2 \RR 3 3          \right. \nonumber\\ &&\left.      +8\mu_1\mu_2\mu_3^2r^4\DD 2 1 ^2          +2\mu_1\mu_2\pp (   8\mu_1\mu_2\mu_3r^2\RR 3 3 +8\mu_1^2\mu_3^2r^2\DD 2 3 +2\mu_1^2\mu_2\mu_3r^2\RR 2 3       \right. \nonumber\\ &&\left.      -3\mu_1^2\mu_3^2r^2\RR 3 3        +3\mu_1\mu_2r^4 \RR 2 3 +\mu_2\mu_3 r^4\RR 1 2 +\mu_2\mu_3^2 r^4\DD 2 1 -4\mu_1\mu_3 r^4\RR 3 3      \right. \nonumber\\ &&\left.  +\mu_1\mu_3^3r^4         +\mu_1\mu_2^3\mu_3^2 r^2         +3\mu_1\mu_2\mu_3^3\DD 2 3 \RR 1 3     )              + \mu_1^2\mu_2\mu_3\pp^2 (   4\mu_2\RR 1 3 \RR 3 3     \right. \nonumber\\ &&\left.   +\mu_1\mu_3^2\RR 2 3     +3\mu_2r^2\RR 2 3     )           \right]\\
\nonumber\\
N^{(2)}_{\omega}&=&  \RR 1 3 \left[      8\mu_1^3\mu_2\mu_3^2 r^4 -8\mu_1\mu_2^2\mu_3^3r^4    +16\mu_1\mu_2^3\mu_3^3\RR 1 3 -16\mu_1^2\mu_2^2\mu_3 r^2\RR 1 3  \right. \nonumber\\ &&\left.         -4\mu_1^3\mu_2^2\mu_3^2r^2\RR 3 3      +    2\mu_1\mu_2^4\mu_3^2r^2\RR 3 3 +2\mu_1^2\mu_2^3\mu_3^2r^2\RR 3 3 - 8\mu_2^3\mu_3^2r^4\DD 2 3      \right. \nonumber\\ &&\left.          +8\mu_1\mu_2^2\mu_3r^6 \DD 2 1  + 6\mu_1\mu_2r^8\DD 2 1 -16\mu_1^2\mu_2^2\mu_3^3r^2\DD 2 1 +8\mu_1\mu_2^2\mu_3^4r^2\DD 2 1      \right. \nonumber\\ &&\left.          -6\mu_1\mu_2^2\mu_3^5r^2\DD 2 1   -8\mu_1\mu_2\mu_3^4r^4\DD 2 1  -8\mu_1\mu_2\mu_3r^4\RR 2 3 \DD 2 1 -2\mu_1\mu_3 r^6\RR 3 3 \DD 2 1      \right. \nonumber\\ &&\left.        -4\mu_2\mu_3r^4\RR 3 3 \RR 2 3 \DD 2 1 +4\mu_1^2\mu_2\mu_3r^4\RR 3 3 \DD 2 1 -2\mu_1\mu_2\mu_3^2r^4\tRR 2 3 \DD 2 1   +8\mu_2^3\mu_3^2 r^4\DD 2 1 \right. \nonumber\\ &&\left.                    +2\mu_2\pp (   \mu_1^3\mu_2^2\mu_3^2r^2 +\mu_1^2\mu_2^2\mu_3^3r^2  - 2\mu_1^2\mu_3r^6   +3\mu_1^2\mu_2^2\mu_3^4\DD 2 3 -8\mu_1\mu_2^2\mu_3^2r^2\DD 1 3           \right. \nonumber\\ &&\left.           -8\mu_1^2\mu_2\mu_3r^2\RR 3 3 +3\mu_1^2\mu_2r^4\RR 2 3 +2\mu_1^2\mu_2^2\mu_3 r^2\RR 2 3  -2\mu_1^2\mu_2\mu_3^2 r^2\RR 3 3            \right. \nonumber\\ &&\left.              -2\mu_1\mu_2\mu_3^3r^2\RR 1 3 -2\mu_1^2\mu_3^3r^2 \RR 2 3  +\mu_1^3\mu_2\mu_3r^2\RR 3 3 +\mu_2^2\mu_3^3r^2\RR 1 2       \right. \nonumber\\ &&\left.        -\mu_1\mu_2\mu_3 r^4\RR 2 3     )          +\mu_1\mu_2^2\mu_3\pp^2 (4\mu_1\RR 2 3 \RR 3 3 +\mu_1r^2\RR 1 3         +3\mu_2\mu_3^2\RR 1 3)              \right] 
\eea
\bea
N^{(3)}_{\omega}&=&   \RR 1 3 ^2 \left[       16\mu_1\mu_2^3\mu_3^3-16\mu_1^2\mu_2^2\mu_3 r^2 + 2\mu_1\mu_2r^2\DD 2 1 \RR 2 3 \RR 3 3 -2\mu_2\mu_3 r^2\DD 2 1 \RR 2 3 \RR 3 3   \right. \nonumber\\ &&\left.    -16\mu_1\mu_2^2\mu_3^2r^2\DD 2 1  + 8\mu_2^2\mu_3^3r^2\DD 2 1-8\mu_1\mu_2\mu_3r^4\DD 2 1  +8\mu_1\mu_2\mu_3r^2\DD 2 1\RR 2 1      \right. \nonumber\\ &&\left.           +2\mu_2\mu_3r^4 \DD 2 1\RR 2 1  +\mu_2^2\mu_3^2\RR 3 3 \DD 2 1\RR 2 1 -\mu_1\mu_2r^2 \DD 2 1\RR 2 1\RR 3 3-\mu_1\mu_3r^2 \DD 2 1\RR 2 1^2       \right. \nonumber\\ &&\left.          +\mu_2\mu_3^3 \DD 2 1\RR 2 1\tRR 2 1       -\mu_1\mu_2^2\mu_3\pp^2(\mu_2\mu_3^2+3\mu_1r^2 )                     \right]   \\
\nonumber\\
M^{(0)}_{\omega}&=&32\mu_1^2\mu_2\mu_3\RR 2 3 ^2(\mu_1\mu_3^2+\mu_2r^2)\\
\nonumber\\
M^{(1)}_{\omega}&=&64\mu_1^2\mu_2^2\mu_3\RR 1 3 \RR 2 3 \RR 3 3\\
\nonumber\\
M^{(2)}_{\omega}&=&32\mu_1\mu_2^2\mu_3\RR 1 3 ^2(\mu_2\mu_3^2+\mu_1r^2)
\eea

\end{appendix}

\end{document}